\documentclass[12pt,a4paper]{article}
\usepackage{textcomp}
\usepackage{fix-cm}

\usepackage[utf8]{inputenc}
\usepackage[a4paper, total={6in, 8in}]{geometry}
\usepackage[onehalfspacing]{setspace}
\usepackage{graphicx}
\usepackage{amsmath, amssymb}
\usepackage{hyperref}
\usepackage[longnamesfirst, authoryear]{natbib}
\usepackage{cleveref}
\usepackage{sansmath}
\setcitestyle{authoryear,open={(},close={)}}
\usepackage{subcaption}
\usepackage{lmodern}
\usepackage{float}
\usepackage{tabularx}
\usepackage{booktabs}
\usepackage[utf8]{inputenc}
\usepackage{caption}
\usepackage{graphicx} 
\usepackage[table]{xcolor}  % For cell colors
\usepackage{colortbl}

\usepackage{array}
\usepackage{caption}
\usepackage{subcaption}
\usepackage{graphicx}       % For \resizebox
\usepackage{threeparttable} % For table footnotesformatting
\usepackage{ragged2e}  
\newcolumntype{Y}{>{\RaggedRight\arraybackslash}X}

\title{An Agent-Based Model for Migration Decision-Making
Under Higher Frequency of Extreme Climate Events}
\author{
  Valentina Antonaccio Guedes\thanks{School of Geography and the Environment. University of Oxford. Email: valentina.antonaccioguedes@ouce.ox.ac.uk }
  \and
  Rafael Prieto-Curiel\thanks{Complexity Science Hub. Vienna. Email: prieto-curiel@csh.ac.at}
}
\date{} 

\usepackage{etoolbox}
\patchcmd{\em}{\itshape}{\emph}{}{}
\begin{document}

% ---------- Title page ----------
\maketitle
%\noindent\textbf{Draft version. Do not distribute or share without the author's permission.}

% ---------- Abstract ----------
\begin{abstract}
This paper develops an agent-based model of climate-related human migration that links repeated environmental shocks to individual migration decision-making through the joint evolution of perceived risk, aspirations to migrate, and migration capability. Building on the aspirations-capabilities framework, the model represents migration as an emergent outcome of two opposing dynamics: shocks increase perceived risk and raise aspirations to move, while simultaneously eroding wealth and reducing the capability to do so. This interaction generates non-linear mobility patterns, including immobility under rising climate stress. Agents are embedded in a spatially heterogeneous shock environment, and their migration capability is further shaped by local conditions and diaspora support. Results show that more frequent shocks can initially increase migration pressure but eventually trap vulnerable agents in place as resources decline. Climate shocks amplify existing income inequalities in mobility outcomes, with lower-income agents more likely to become trapped or remain acquiescently immobile, while higher-income agents retain greater flexibility to migrate or choose voluntary immobility. %By capturing heterogeneous responses and feedbacks at the micro level, the paper offers a process-based perspective on climate mobility and provides a foundation for future work on social networks, institutional support, and policy interventions.

\end{abstract}

\section{Introduction}
%%%%%%%%%%%%%%%%%%%%%%%%%%%%%%%%%%%%%%%%%%
%% 1. INTRODUCTION
%%%%%%%%%%%%%%%%%%%%%%%%%%%%%%%%%%%%%%%%%%

% 1) Motivation: Explain why modeling migration under climate change is important. Climate change alters environmental conditions, migration is one adaptive response and aggegate migration patters emerge from decentralised decisions. Responses are heterogenous and often contradictory. 
{
Climate change affects people’s livelihoods both directly, through environmental impacts, and indirectly, via economic, political, and social channels. Migration is one of the key strategies households use to cope with or adapt to these effects \citep{knight_coming_2022}. As droughts and floods become more frequent and intense, internal migration is expected to become an increasingly central policy concern over the coming decades, particularly in developing countries \citep{rigaudGroundswell2018}. Here, we understand internal migration as people’s movement within national borders, including rural-urban, urban-rural and inter-urban flows \citep{foresightMigrationGlobalEnvironmental2011}. Despite this growing relevance, there is little consensus on the magnitude or direction of the relationship between climate change and migration \citep{kaczan_impact_2020}. This reflects that migration is only one of several possible adaptation strategies and is shaped by multiple constraints and frictions \citep{thiede_climate_2022}. Moving requires not only the desire to leave, but also the capability to do so \citep{de_haas_theory_2021}. For example, evidence shows that when farmers are exposed to higher-than-normal temperatures, their primary source of income declines, leaving them trapped \citep{cattaneoMigrationResponseIncreasing2016}. This challenges the assumption that climate change will lead to large-scale displacement and instead highlights the conditions under which populations may become trapped. 
}

% 2) Other methods to study migration and climate change. 
% Gravity modeles operate at aggregate scale, utility-based models assume frictionless optimization and empirical approaches identify correlations but not mechanisms. 
% These approaches do not explicitly integrate endogenous risk perception, resource dynamics and threshold-based decision rules wihtin a unified dynamic system. 
{
Although migration flows are large in absolute terms, most people do not move. As of 2020, approximately 96\% of the global population lived in their country of birth, and even internal migrants represent a minority share of the population \citep{united_nations_department_of_economic_and_social_affairs_emigrants_2024}. Yet, projections suggest that climate-related mobility may increase substantially in the coming decades, with estimates reaching hundreds of millions of internal climate migrants by mid-century \citep{kaczan_impact_2020}. Large-scale assessments, such as the World Bank’s Groundswell report, generate these projections using gravity-type models grounded in push-pull theory \citep{lee_theory_1966, rigaudGroundswell2018}. In such frameworks, migration flows are primarily driven by differences in location-specific attractiveness and distance-related costs, making them particularly well-suited to explain labour migration, where access to employment is a central driver. While useful for aggregate forecasting, these approaches abstract from the micro-level mechanisms through which environmental shocks shape risk perception, resource constraints, and individual decision processes \citep{klabunde_decision-making_2016}. As a result, they provide limited insight into how heterogeneous responses and potentially non-linear mobility dynamics emerge from individual adaptation under climate stress.
}

{
Efforts to identify the mechanisms underlying climate-related migration span both empirical and computational approaches. Experimental studies have examined how information provision, financial assistance, and job-search support influence mobility decisions, including specific barriers to movement \citep{mckenzie_fears_2024}. However, these approaches typically focus on local treatment effects and do not capture the broader system-level dynamics that emerge under repeated climate stress. Computational models have sought to address this limitation. Existing agent-based models of climate-related mobility often conceptualise migration as a household risk-management or income-smoothing strategy, grounded in the New Economics of Labour Migration framework, in which one member migrates in search of work and sends remittances back to their household \citep{starkNewEconomicsLabor1985}. These frameworks typically incorporate heterogeneous agents, bounded rationality, and social interactions to capture how environmental variability and policy interventions shape livelihood and migration decisions \citep{choquette-levy_impact_2019}. While this approach provides important insights into the role of resource constraints and institutional support, it generally does not explicitly account for the influence of diaspora networks, individual agency or even risk perception in shaping mobility decisions. As a result, they offer limited insight into how these factors jointly determine both the likelihood of migration and the emergence of constrained or immobile populations under climate stress.
}

% 4) What does the theory has to say about migration.  

{
In contrast, the aspirations-capabilities framework conceptualises migration as the outcome of the interaction between individuals’ aspirations to migrate and their capabilities to do so, both of which are shaped by broader structural opportunity contexts \citep{de_haas_theory_2021}. Migrating and staying are thus complementary expressions of agency, as they lie at the intersection between the individual actor and the context within which they live \citep{kingTheoriesTypologiesMigration2013}
}

{
Within this perspective, environmental migration can be understood along a continuum from voluntary to forced mobility, sometimes described as proactive versus reactive movement \citep{vinke_migration_2020}. Forced migration occurs when individuals have no realistic option but to leave, as in contexts of extreme violence, persecution, or acute environmental threats \citep{de_haas_theory_2021}. In climate mobility research, displacement typically refers to situations in which hazard exposure and risks to life or physical integrity dominate decision-making, whereas more voluntary forms of migration arise when environmental change interacts with economic or social aspirations, and individuals retain meaningful choice \citep{desherbininMigrationTheoryClimate2022}. Agency in this framework is defined as the real freedom to choose between moving and staying. Under low to moderate environmental stress, individuals with sufficient resources can exercise greater control over this choice. As stress intensifies, however, capabilities contract and decisions become increasingly constrained, rendering both movement (displacement) and non-movement (entrapment) more involuntary outcomes \citep{romdhaneDisplacementEntrapmentClimateinduced2025}.
}

%  5) Objective : State clearly that this is a theoretical ABM exploring mechanisms and behavioural interactions. 
{
In this paper, we develop an agent-based model of migration under climate-related shocks in which heterogeneous agents are exposed to stochastic extreme events whose frequency increases over time. Rather than assuming shocks become more severe, we examine how their increasing frequency reshapes decision-making dynamics. Repeated exposure influences perceived risk, thereby shaping the desire to migrate, while simultaneously affecting both material resources and social capital, which together determine migration capabilities. Mobility occurs only when both aspiration and capability exceed a specific threshold. By combining evolving risk perception with resource- and network-dependent capability constraints, the model captures how increasing shock frequency can generate heterogeneous mobility regimes, including migration, voluntary immobility, and entrapment.
}

% 6) Contribution summary / What the model does. 
{
Our results highlight a clear asymmetry between desire and capability in shaping mobility outcomes. Across climate scenarios, immobility remains the dominant outcome, but the aggregate distribution is no longer stable: increasing climate pressure produces a systematic shift away from acquiescent and voluntary immobility and towards both migration and trapped immobility. This suggests that stronger climate intensity does not generate a uniform mobility response. Instead, it polarises outcomes: some agents are pushed across the migration threshold, while others experience rising desire without sufficient capability to move. Disaggregating by wealth shows that individuals in the lower quintiles are disproportionately acquiescent or trapped, indicating limited capability despite potential exposure, whereas richer groups are more likely to migrate or choose voluntary immobility. Under the shocked climate, this inequality sharpens: trapping increases among lower-wealth groups, while migration rises mainly among those with sufficient resources. Overall, the model shows that climate shocks amplify existing inequalities by expanding mobility for some while deepening constrained immobility for others.
}

\section{The Agent-Based Model}

\subsection{Model overview}

{
We develop an agent-based model in which migration decision-making emerges from coupled dynamical processes governing (i) perceived risk, (ii) desire to migrate, and (iii) capability to migrate. The desire-capability mechanism follows the Aspirations and Capabilities framework, operationalised here as a system of dynamical equations updated at discrete time steps \citep{de_haas_theory_2021}. Each time step corresponds to a week, and the model is simulated over a 100-year horizon. It contains two key feedback structures: (i) repeated shocks increase perceived risk, raising the desire to migrate; (ii) repeated shocks erode wealth, reducing migration capability. The interaction between these opposing forces generates non-linear mobility outcomes. We focus on proactive migration decisions, in which individuals retain agency to decide whether to move or stay, abstracting from displacement scenarios in which relocation is externally imposed \citep{vinke_migration_2020}. The model examines how the increasing frequency of extreme climate events reshapes the joint dynamics of aspiration and capability, and under what conditions this produces migration, voluntary immobility, entrapment, or acquiescence.
}

\subsection{Environment and extreme climate events}

{
The environment is a grid of cells, $\mathcal{G}$, each representing a different location. 
\[
\mathcal{G} = \{1,\dots,G\}\times\{1,\dots,G\}, \qquad |\mathcal{G}| = G^2.
\] 
There are $N$ agents indexed by $i\in\{1,\dots,N\}$. Each agent occupies a cell
\[
x_i(\tau) \in \mathcal{G},
\]
potentially, with more than one agent per cell. Cells evolve independently; no spatial spillovers or spatial autocorrelation are modelled. Extreme climate events (e.g. droughts, floods) occur stochastically in each cell $c$, in which the rate increases linearly with time, so they become progressively more frequent. This is characterised by an initial Poisson event-arrival rate $\lambda_{0,c}$, drawn from a uniform distribution. Using the parameter $\alpha_c$, we control the local strength of climate intensification in cell $c$, that is, how much the arrival of these extreme events increases for each of these cells. As $\alpha_c$ increases, the model generates competing forces: higher shock frequency raises aspiration via perceived risk, while simultaneously reducing capability through wealth erosion. The relative strength of these processes determines whether migration increases, stalls, or transitions into expanding immobility traps. The arrival of events evolves following Equation \ref{eq:climateshocks}. When $\alpha_c =0$, extreme climate events happen with the baseline rate, $\lambda_{0,c}$ (Figure \ref{fig:climatechange}). This is not a critical assumption given the 100-year simulation horizon, and the IPCC finding that many extremes scale quasi-linearly with global warming levels, we approximate the arrival rate of climate shocks as increasing linearly over time from a baseline rate to scenario-specific end-of-horizon rates \citep{intergovernmentalpanelonclimatechangeipccClimateChange20212023}. This linearization is a simplifying assumption and should be interpreted as a first-order approximation rather than a literal description of climate dynamics.

\begin{equation}
    \lambda_c(\tau) = \lambda_{0,c} + \alpha_c \tau \text{, with} 
\qquad \tau = t\,\Delta t.
\label{eq:climateshocks}
\end{equation}
The model isolates the effect of increasing event frequency while holding event intensity constant, allowing us to examine how changes in arrival rates alone reshape decision dynamics. Continuous-time equations are implemented numerically using discrete-time updates with step size $\Delta t$.

\begin{figure}[H]
    \centering
    
    \begin{subfigure}[t]{0.48\textwidth}
        \centering
        \includegraphics[width=\linewidth, trim=4cm 0 4cm 0, clip]{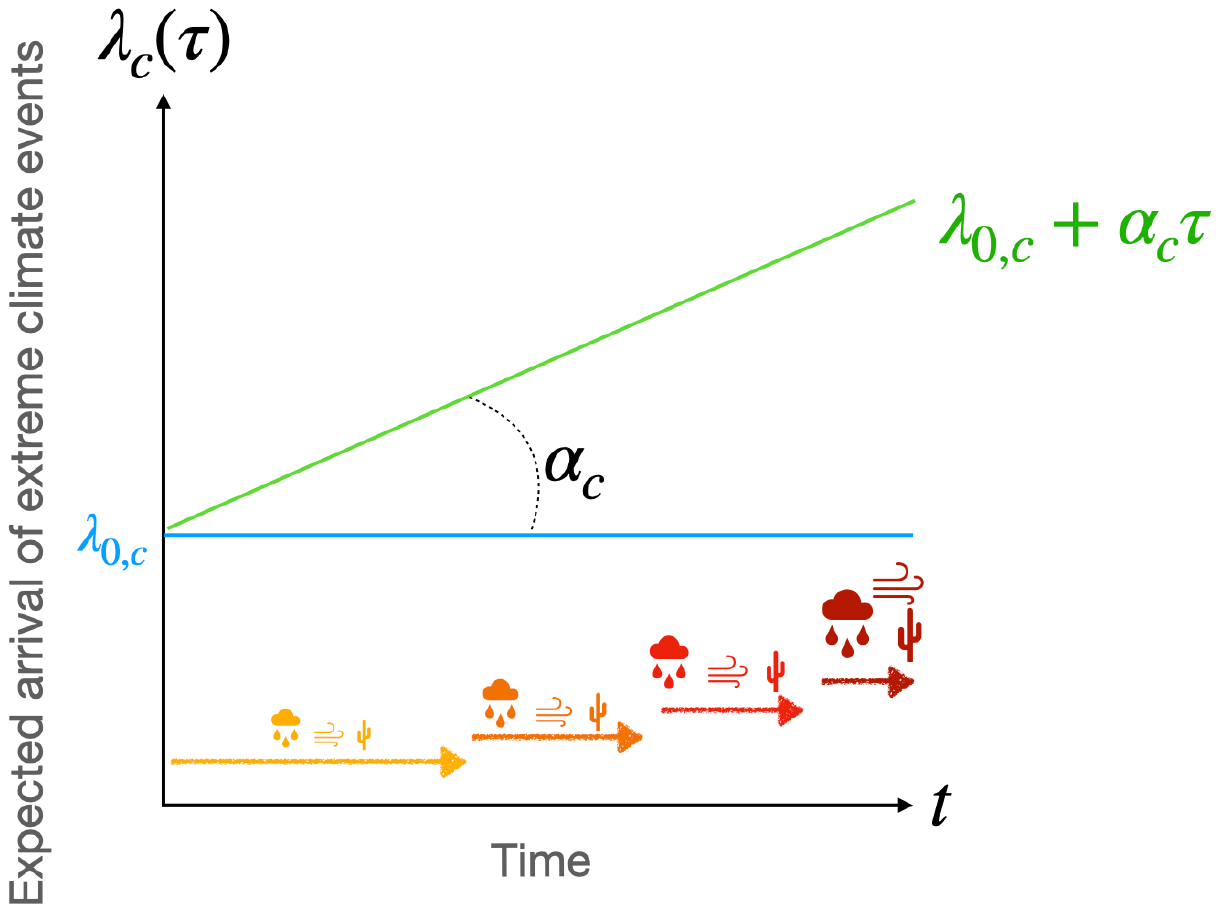}
        \caption{Arrival of climate change events}
        \label{fig:climatechange}
    \end{subfigure}
    \hfill
    \begin{subfigure}[t]{0.48\textwidth}
        \centering
        \includegraphics[width=\linewidth, trim=4cm 1cm 4cm 0.5cm, clip]{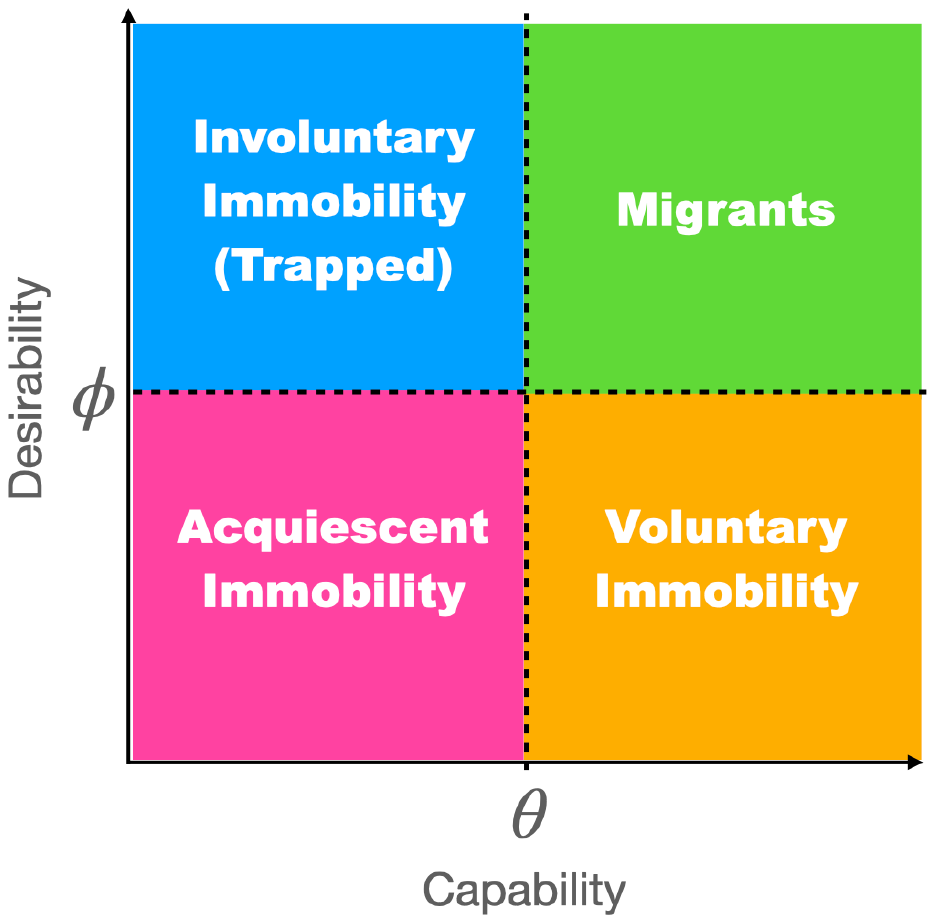}
        \caption{Decision-making mechanisms}
        \label{fig:Desire-Capability}
        
       % {\footnotesize \textit{Note.} Migration is modelled here with the \cite{de_haas_theory_2021, carling_migration_2002, schewelUnderstandingAspirationStay2015, schewel_understanding_2020, leerkes_what_2017}  model}
    \end{subfigure}
    
    \caption{Conceptual structure of the model: (a) climate-driven shocks and (b) the desire-capability decision framework.}
    
    \label{fig:conceptual_panels}
\end{figure}

}
\subsection{Agents and perceived risk}

%%%% perception of risks
{
Agents are initially distributed randomly across the grid. Each cell has a count of shocks that have happened: $K_c(t)$ that follows a Poisson distribution with parameter $\lambda_c(\tau_t)$. Each agent, $i$ at current location $x_i(t)$  updates their perceived risk $P_i$ using the shocks in the current cell $n_i(t) = K_{x_i(t)}(t)$. Each agent $i$ experiences a time-varying perception of risk, $P_i(\tau)$,
\begin{equation}
\frac{d P_i}{d\tau} = -\gamma_iP_i(\tau) + \eta_i \xi_i(\tau),
\label{eq:perception}
\end{equation}

while $n_i(t)$ is the number of shocks in a discrete time step, $\xi_i(\tau)$ is the instantaneous shock exposure signal in continuous time, $\eta_i$ scales the impact of experienced events on perception, and $\gamma_i > 0$ controls memory decay. Thus, risk perception decreases exponentially in the absence of shocks and increases with exposure to events. This specification yields an exponentially decaying memory of past shocks. 
}

\subsection{Desire to migrate}

%%%%
{
Following the Aspirations and Capabilities framework \citep{de_haas_theory_2021}, perceived risk shapes the agent’s desire to migrate, $D_i(\tau)$. We model desire as a relaxation process toward perceived risk:
\begin{equation}
\frac{dD_i}{d\tau} = r(P_i(\tau) - D_i(\tau)),
\label{eq:desire_ode}
\end{equation}
where $r>0$ is an adjustment speed. This specification is analogous to Newton’s law of cooling: perception $P_i(\tau)$ acts as the time-varying ``ambient'' level, and $D_i(\tau)$ adjusts toward it. Desire does not respond instantaneously to perception but adjusts gradually. 

Agents start at equilibrium because they are considered to be in an environment familiar to them. The arrival of climate shocks is not seen as a surprise but as something that occurs at a known frequency. For example, every winter there is a flood. This is part of the ``Business As Usual'' scenario. In equilibrium, we have that 
\begin{equation*}
    \frac{dD_i}{d\tau} = 0,  
\end{equation*}
so, 
\begin{equation*}
    P_i^* = D_i^*.
\end{equation*}

The risk perception in equilibrium is given by:
\begin{equation*}
    P_i^*= \frac{\eta_i \lambda_{0,c}}{\gamma_i}.
\end{equation*}

Upon migration, agents update their perceived risk as a weighted average of their prior perception and the equilibrium risk of the destination cell:  
\begin{equation}
    P_i(\tau+\Delta t)= \kappa_i P_i(\tau) + (1-\kappa_i) P_i^*,
\end{equation}
where $P_i(\tau)$ is their old risk perception, $P_i^*$ is their new risk perception, which is in equilibrium, using the expected arrival of events $\lambda_{0,c}$ specific to their new cell. $\kappa_i$ is an agent-specific adjustment parameter. That is, how much the agent trusts their previous experience and how quickly they adapt to their new place. 

}

\subsection{Capability to migrate: wealth and social capital}

An agent’s migration capability, $C_i(\tau)$, is determined by wealth $W_i(\tau)$ and social capital $S_i(\tau)$:
\begin{equation}
C_i(\tau) = \psi_W W_i(\tau) + \psi_S S_i(\tau),
\label{eq:capability}
\end{equation}
where $\psi_W$ and $\psi_S$ weight the relative contributions of economic resources and network resources.

Wealth evolves in discrete time and is negatively affected by shocks. Continuous-time equations are numerically approximated via forward Euler discretisation with time step $\Delta t$. Let $n_i(\tau)$ denote the shock count experienced by agent $i$ during time step $t$. Then
\[
W_i(\tau+\Delta t)=
\begin{cases}
(1-\delta_{W,i})\,W_i(\tau), & \text{if } n_i(\tau) > 0 \text{, and}\\[6pt]
\alpha_{W,i}\,W_i(\tau) + \big(1-\alpha_{W,i}\big)\,W_{i,0}, & \text{if } n_i(\tau)=0,
\end{cases}
\]
where $W_{0,i}$ is baseline wealth and $\alpha_W\in(0,1)$ controls the speed of recovery towards baseline in shock-free periods. This formulation models mean reversion toward the baseline wealth level.

Social capital $S_i(\tau)$ captures access to local and diaspora ties. Specifically, it combines (i) the share of individuals known within the agent’s current cell and (ii) ties to residents in cells that have previously received migrants originating from the agent’s cell. This captures both local embeddedness and migration-network effects. Full construction details and parametrisation are provided in the Appendix. 

\subsection{Migration decision rule and mobility states}

At each time step, agents are assigned to one of four mutually exclusive states based on their location in the desire-capability plane (Figure~\ref{fig:Desire-Capability}). Two fixed thresholds, desire $\phi$ and capability $\theta$, divide the plane into four quadrants. The thresholds act as binary gates: when an agent’s desire crosses $\phi$ (from $D_a(\tau)<\phi$ to $D_a(\tau)\ge \phi$), it is reclassified from “low desire” to “high desire”; likewise, when capability crosses $\theta$ (from $C_a(\tau)<\theta$ to $C_a(\tau)\ge \theta$), it is reclassified from “low capability” to “high capability.” Thus, any crossing of either threshold immediately moves the agent into a different category at that time step, which in turn changes the decision rule for whether the agent migrates. The model therefore contains four interacting timescales: shock arrival, perception decay, aspiration adjustment, and economic recovery.

Migration is modelled as requiring both aspiration and capability, consistent with the aspirations-capabilities framework. In this sense, the two dimensions are complementary rather than substitutable: strong desire alone cannot generate migration in the absence of sufficient capability. This abstraction captures a threshold condition for mobility, while recognising that in reality individuals may shift to alternative migration modes with lower capability requirements. This is especially the case for individuals who may have a very strong desire while their capability is near the threshold \citep{schewel_understanding_2020, de_haas_theory_2021, carling_migration_2002}. 

\begin{itemize}
\item If $D_i(\tau)<\phi$ and $C_i(\tau)<\theta$, the agent remains in place in a state of \textit{acquiescent immobility} \citep{de_haas_theory_2021, schewel_understanding_2020, schewelUnderstandingAspirationStay2015}.
\item If $D_i(\tau)\ge \phi$ but $C_i(\tau)<\theta$, the agent is \textit{involuntarily immobile} (``trapped'') \citep{de_haas_theory_2021, carling_migration_2002}.
\item If $D_i(\tau)<\phi$ but $C_i(\tau)\ge \theta$, the agent is \textit{voluntarily immobile}: it could migrate but does not desire to do so, potentially due to alternative adaptation options.
\item If $D_i(\tau)\ge \phi$ and $C_i(\tau)\ge \theta$, the agent migrates.
\end{itemize}
\section{Results}
\subsection{Climate Change, Migration and Wealth Inequality} 

% Intro 

{The model captures how increasing climate pressure can generate heterogeneous and non-linear mobility outcomes driven by individual desire and capabilities. At each time step, agents are classified according to whether their desire and capability exceed fixed thresholds. Migration occurs only when both are sufficiently high; otherwise, agents remain in place as voluntarily immobile, acquiescent, or trapped. We compare two climate scenarios based on the IPCC's $6^{th}$ Assessment Report's projected changes in the intensity and frequency of hot temperature extremes and extreme precipitation over land, as well as agricultural and ecological droughts in drying regions \citep{ipccSummaryPolicymakers2023}. First, we have Business As Usual (BAU), which is our current scenario of $1^oC$ temperature. It represents a stationary climate risk environment in which the frequency of hazardous events remains constant over time, as if the climate would not change any further from what it already is. It is meant to provide a reference case against the ``shocked'' scenario, in which the arrival of extreme events changes linearly with warming, according to warming-level scenarios of $1.5^oC$, $2^oC$, and $4^oC$ global temperature increase. The remaining parameters are chosen so that our results do not depend on a single value. All results are robust across a wide range of parameters in the sensitivity analysis, using both one-at-a-time and variance-based approaches. 
}

% Main result. What is happening? 
{
While immobility remains the dominant outcome across all scenarios, increasing climate pressure produces a clear redistribution of mobility states. As warming increases, both acquiescent and voluntary immobility decline, while trapped immobility and migration increase. This suggests that climate change does not simply generate immobility under constraint; it also pushes a growing share of agents across the migration threshold. However, the increase in migration occurs alongside a rise in trapped populations, indicating that climate pressure polarises mobility outcomes: some agents become more likely to move, while others experience a growing desire without sufficient capability to act. In this sense, stronger climate pressure expands mobility for some agents but deepens constrained immobility for others (Figure \ref{fig:main_results}).
}

% Why are we getting these results? 
{
These results are driven by the asymmetric effects of shocks on desire and capability. Repeated shocks increase the desire to migrate by raising perceived risk, but they can also erode wealth, thereby reducing capability. For agents with sufficient resources, rising perceived risk can prompt migration. For poorer or more vulnerable agents, however, the same shocks increase desire without generating the capability required to move, producing trapped immobility. The aggregate result is therefore not a uniform shift towards migration or immobility, but a divergence between those able to convert climate pressure into mobility and those whose mobility becomes increasingly constrained. Therefore, we can conclude that there is an unequal adaptive capacity. 

\begin{figure}[H] \centering \includegraphics[width=18cm,trim=0 0cm 0 0cm,clip]{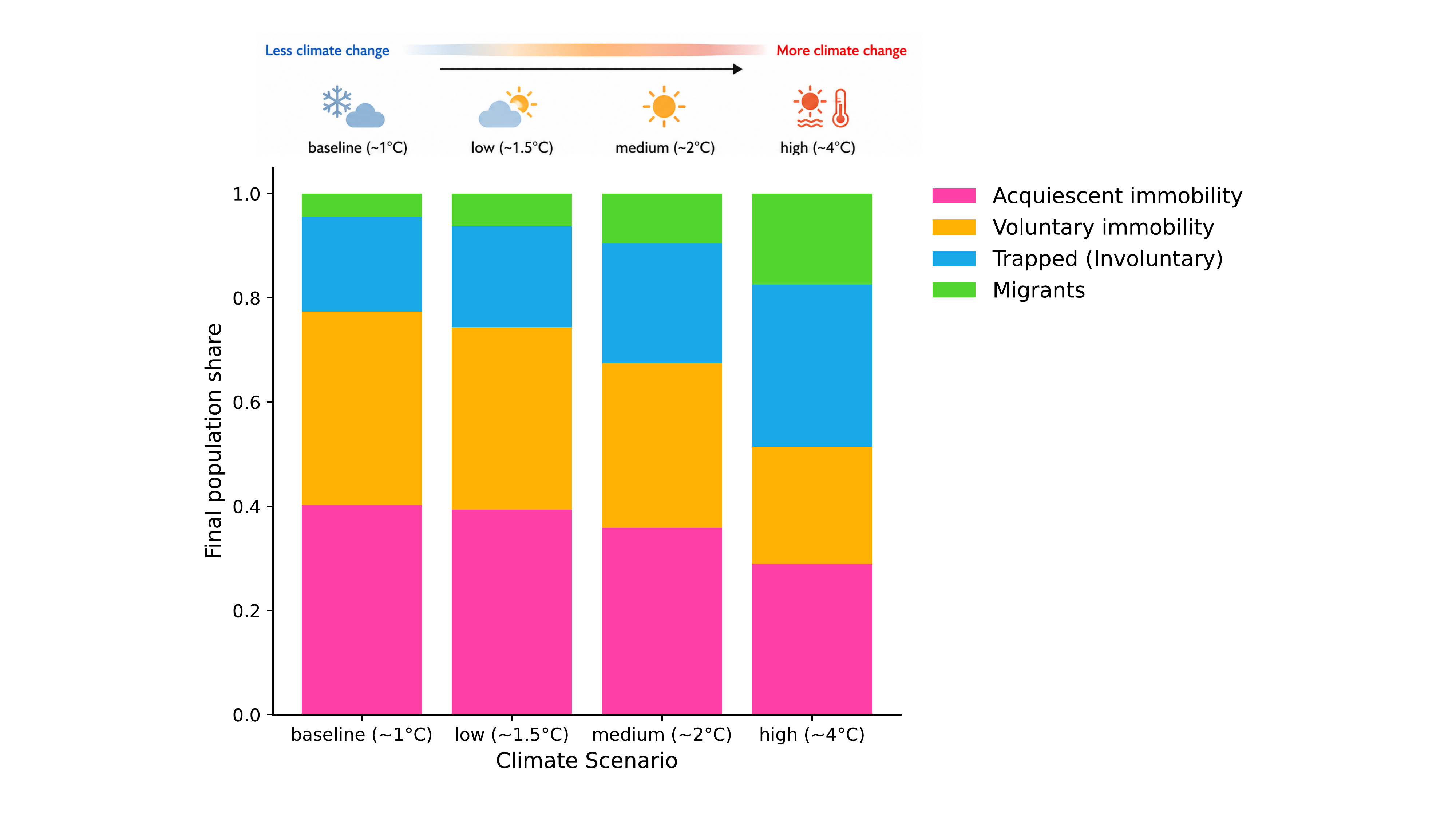} \caption{Aggregate distribution of mobility states across climate scenarios. Increasing climate pressure reduces both acquiescent and voluntary immobility, while increasing both trapped immobility and migration. This indicates that stronger climate risk polarises mobility outcomes: some agents become more likely to migrate, while others become increasingly constrained. The climate scenarios are informed by the IPCC $6^{th}$ Assessment Report projections of changes in the intensity and frequency of climate extremes, including hot temperature extremes, extreme precipitation, and droughts \citep{ipccSummaryPolicymakers2023}} \label{fig:main_results} 
\end{figure} 

} 

\begin{figure}[H] 
\centering 
\includegraphics[width=13cm]{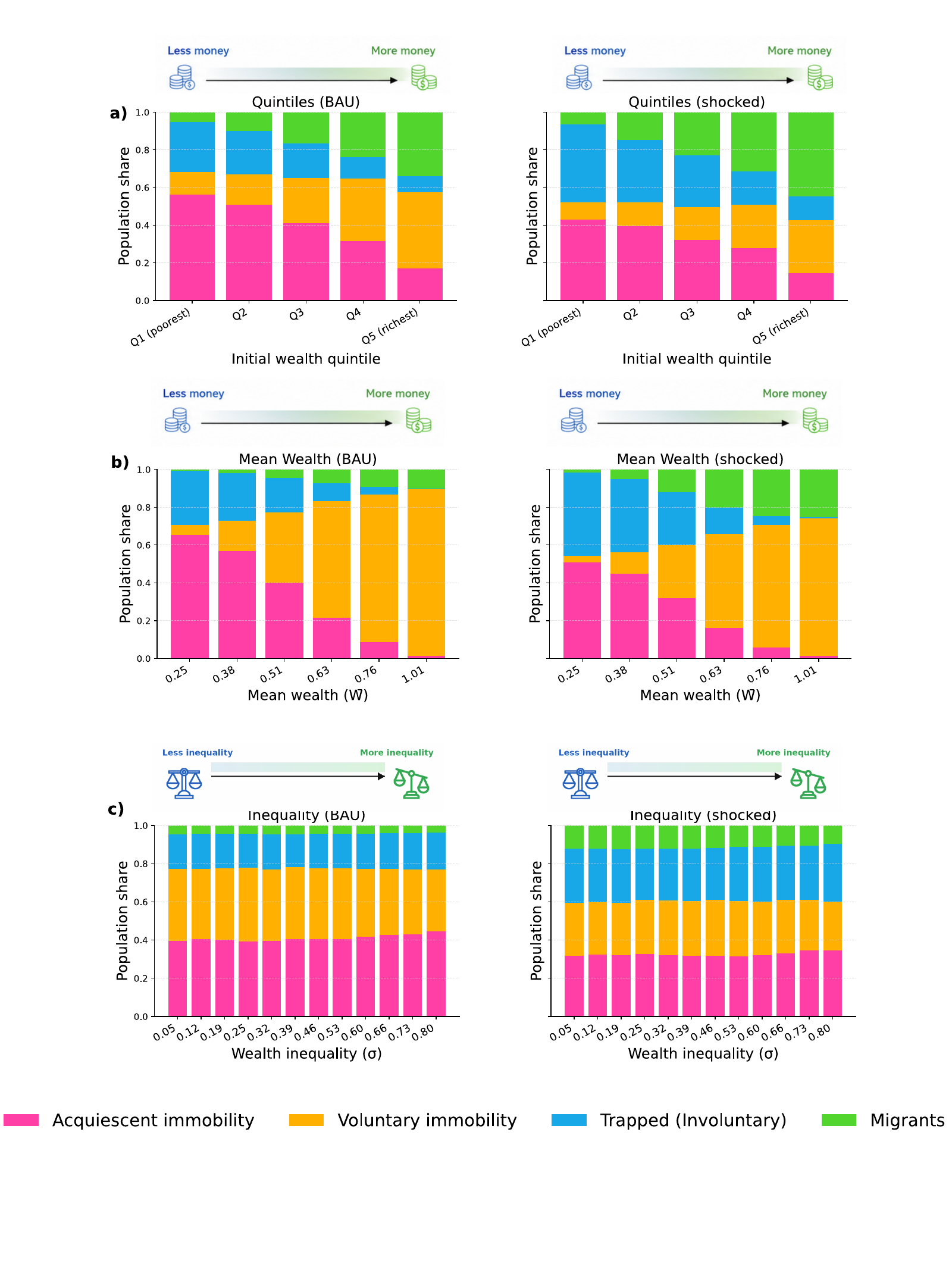} 
% \vspace{0.3cm}  
\caption{Mobility outcomes across wealth distributions and inequality under baseline and climate scenarios. Panels (a-c) show the distribution of agents across mobility states (acquiescent, voluntary immobile, trapped and migrants) as a function of (a) initial wealth quintiles, (b) mean wealth, and (c) wealth inequality. In each panel, we compare business-as-usual (left) with a shocked-climate scenario (right).} 
\label{fig:wealth_inequality} 
\end{figure}

%Who are the agents that are trapped? Inequality story % wealth distribution 
{
The counterpoising forces between capability degradation and an increase in desire are amplified by wealth inequality (Figure \ref{fig:wealth_inequality}, panel a). Poorer agents (in the lower quintiles) are disproportionately concentrated in acquiescent or trapped states, while wealthier agents are more likely to achieve voluntary immobility or migration states. Under climate shocks, this stratification intensifies. When shocks degrade agents' capabilities, lower-wealth groups fall further into trapped categories, while wealthier groups adapt in other ways, including migration.
} 

% Are middle income societies more mobile than very unequal ones? How does mobility outcomes change with income? 
{
Richer people are not more migratory than poorer ones (Figure \ref{fig:wealth_inequality}, panel b). The wealthier a society becomes, the more agents shift from acquiescent and trapped states to voluntary immobility. So, as choice expands, people do not necessarily move more. In fact, migration remains quite constant. What we are seeing instead is that intermediate wealth levels represent a critical transition zone in which agents are both willing and able to move. Under climate shocks, this transition is delayed, with more agents remaining in constrained states at comparable wealth levels. These results indicate that mobility is not at its highest point in the poorest or richest contexts, but where capability constraints are sufficiently relaxed while incentives to move remain salient.
} 

% What happens with high inequality or low inequality? How does this affect the mobility outcomes? 
{
Finally, when holding mean wealth constant and systematically increasing inequality (the variance of the log-normal distribution of wealth), the model generates a progressively more constrained mobility landscape. Acquiescent immobility increases, voluntary immobility slightly decreases, trapped populations increase, while migration remains a small proportion and quite constant overall (Figure \ref{fig:wealth_inequality}, panel c). 

This pattern emerges because higher inequality redistributes capability unevenly across the population. A larger share of the agents falls into the lower tail of the wealth distribution, reducing their ability to respond to growing climate risk. At the same time, a smaller group retains sufficiently high capability to migrate when shocks occur. As a result, inequality polarises mobility outcomes: high-desire agents increasingly separate into two distinct groups: those with enough resources to move and those who become trapped in place. 

The result therefore suggests that inequality does not simply reduce average mobility; rather, it amplifies divergence in adaptive capacity. Climate stress increasingly translates into involuntary immobility for poorer agents, while migration becomes concentrated among the relatively advantaged. In this sense, inequality acts as a mechanism that converts climate exposure into stratified mobility outcomes, generating trapped populations even when aggregate migration levels change little. 

Comparing business as usual with a shocked environment further reinforces this dynamic. Under repeated climate shocks, both trapped populations and migrant populations increase more steeply as inequality rises. This indicates that climate stress magnifies the distributional consequences of unequal capability: shocks raise migration pressure across the population, but only those with sufficient resources can translate aspiration into movement, while vulnerable agents become immobilised. 
}

% what happens if we vary the arrival of climate events. 
{
An increase in the arrival of extreme climate events has an impact on mobility by changing the frequency $\lambda_{0,c}$ (Figure \ref{fig:wealth_inequality} panel d). Even in the absence of rising climate risk, a higher baseline intensity systematically shifts populations away from voluntary immobility towards more constrained states. For reference, a baseline intensity of $\lambda_{0,c} = 0.08$ corresponds to approximately four extreme climate events per year. As the baseline intensity rises, acquiescent immobility declines, while both trapped populations and migration increase. However, this increase in migration remains modest relative to the growth in constrained states, particularly under the shocked climate scenario. In that case, higher baseline risk amplifies the divergence between desire and capability, accelerating the transition of agents into trapped conditions. These results highlight that it is not only changes in climate trends but also the underlying level of environmental stress that shapes mobility outcomes and reinforces immobility under constraint.
}

\subsection{Risk Sensitivity and Memory Decay}

%%%%
{
Risk perception operates through two distinct channels: persistence ($\gamma$) and intensity ($\eta$). Lower memory decay ($\gamma$) prolongs the influence of past shocks, sustaining elevated risk perceptions, modestly increasing migration, and reducing acquiescent immobility. In contrast, higher shock sensitivity ($\eta$) amplifies the immediate impact of each event, sharply reducing acquiescent immobility but disproportionately increasing involuntary immobility. This occurs because heightened perceived risk raises migration aspirations faster than capabilities can adjust, leading to a growing population of trapped individuals. Together, these results highlight that stronger climate risk perception does not translate linearly into higher migration, but can instead generate immobility traps (Figure \ref{fig:wealth_inequality} panels e and f). 
}
\section{Discussion}
% Interpret insights: What do these patterns suggest about migration under risk? 
{
Current research estimates that by 2050, the number of both internal and international migrants will increase as more people are displaced by the growing intensity and frequency of extreme climate events \citep{rigaudGroundswell2018, rigaud_groundswell_2021, kaczan_impact_2020}. Many of these projections are based on gravity models, which emphasise push-pull factors that drive individuals away from origin locations and attract them to destinations offering greater economic opportunities, services and amenities. However, the Aspirations and Capabilities framework suggests that mobility outcomes may follow more complex patterns. Climate change can exert two competing forces that generate non-linear mobility dynamics: increasing exposure to climate shocks may raise individuals' desire to migrate, while simultaneously eroding their capability to do so \citep{de_haas_theory_2021, internal_displacement_migration_center_idmc_2023}. When capability declines faster than desire increases, individuals may become trapped. The broader mobility landscape also includes acquiescent immobility (low desire and low capability) and voluntary immobility (low desire but sufficient capability). 
}

%  * Relate findings back to theory: Do social networks accelerate or delay migration? Does information availability reduce risk-driven migration?

{
A central insight of the model is that climate-related migration cannot be understood as a direct or proportional response to environmental stress. Instead, migration emerges from the interaction between two opposing dynamics: climate shocks increase perceived risk and strengthen the desire to move, while simultaneously eroding the resources required to do so, which is rebuilt while the agents are not affected by another shock. By modeling these feedback explicitly, the framework generates non-linear mobility outcomes endogenously rather than imposing them through fixed assumptions. Stronger climate pressure therefore does not produce uniform increases in migration. Instead, the system progressively polarises into groups that are able to convert rising risk into mobility and groups that become trapped despite increasing desire to leave. 
}

{
This mechanism helps explain why stronger climate pressure does not translate into a simple or proportional increase in migration. While migration rises across climate scenarios, a substantial share of the adjustment occurs through expanding trapped populations rather than through mobility alone. As climate stress intensifies, acquiescent and voluntary immobility decline, while both migration and trapped immobility increase. The model therefore suggests that climate change reshapes the distribution of mobility outcomes as much as it increases movement itself. Rising climate risk generates growing pressure to move across the population, but unequal capabilities determine whether this pressure results in migration or involuntary immobility. 
}
%%%%
{
The model further demonstrates that wealth inequality is not simply a background condition affecting migration levels, but a structural mechanism that determines how climate risk is translated into mobility outcomes. Holding mean wealth constant while increasing inequality produces sharply different trajectories across agents. Individuals in the lower tail of the wealth distribution increasingly lose the capability to respond to rising climate risk, leading to expanding trapped populations, while wealthier agents retain the ability to migrate. Inequality therefore polarises adaptive responses: climate shocks generate similar pressures across the population, but only some agents are able to transform aspiration into movement.  
}

{
This finding is consistent with a well-established literature showing that migration is shaped by resource constraints. Individuals in lower wealth quintiles are more likely to be either acquiescent immobile or trapped, while wealthier individuals are more likely to be voluntarily immobile or to migrate, reflecting their greater capacity to access a broader set of adaptation options \citep{mckenzie_poverty_2017, dustmann_out-migration_2014}. Under climate shocks, this stratification intensifies, with poorer groups becoming more trapped and wealthier groups more able to translate rising risk into migration.
}

{
This result also highlights an important advantage of the modeling approach. Standard gravity and push-pull frameworks typically treat migration as an aggregate flow responding to average conditions. In contrast, the agent-based framework captures how heterogeneous capabilities generate divergent responses to the same environmental pressures. Aggregate migration can remain relatively constant even while underlying vulnerability and involuntary immobility increase substantially. The model therefore suggests that climate mobility shoudl be understood less as a question of how many people move, and more as a question of who is able to move under conditions of rising risk. 
}

%%%%
{
At the aggregate level, our results suggest that richer societies are not necessarily more migratory than poorer ones. Instead, migration does not increase monotonically with average wealth. This pattern is consistent with the Aspirations and Capabilities framework proposed by \cite{de_haas_theory_2021}, which highlights that rising incomes do not automatically reduce migration. Empirical observations, such as migration from the Todgha Valley to urban centres in Morocco, show that migration can persist despite substantial improvements in income and living standards. As \cite{de_haas_theory_2021} argues, this contradicts the predictions of neoclassical and push-pull models, which would anticipate declining migration as origin conditions improve. Rather, intermediate levels of wealth emerge as a critical transition zone in which individuals are both willing and able to move. Under climate shocks, this transition is delayed, with more individuals remaining in constrained states at comparable wealth levels. These findings indicate that migration is not highest in the poorest or richest contexts, but instead peaks where capability constraints are sufficiently relaxed while incentives to move remain strong. 
}

% Implications: How can policies (e.g. early warning, subsidies) affect adaptive migration? 
{
The model highlights that migration outcomes are jointly shaped by rising risk perceptions and declining mobility resources. This implies that policy interventions should move beyond a narrow focus on migration incentives and instead aim to align aspirations with capabilities. In particular, reducing liquidity constraints through targeted financial support, expanding safe and legal migration pathways, and leveraging diaspora networks can facilitate mobility where it is desired. At the same time, sustained investment in local adaptation is essential for populations that are unable or unwilling to move, helping to prevent the emergence of trapped populations under increasing climate stress.  
}

% Limitations: Model simplicity, theoretical nature, no empirical grounding yet. 
{
This model adopts a deliberately stylised representation of migration decision-making, focusing on the interaction between perceived risk, migration desire and financial capability. As such, several limitations should be noted. First, behavioural complexity is reduced to a parsimonious decision structure, abstracting from social norms and heterogeneous preferences. Second, migration capability is proxied primarily by wealth, omitting institutional, legal and social constraints; in particular, the model assumes that agents who both want and can migrate are able to do so without restriction. Third, the spatial and network structures are simplified, with diaspora effects captured at the aggregate cell level rather than through explicit social ties. 
}

%%%%
{
These modelling choices also shape key results. The emergence of mobility traps is closely linked to the direct erosion of wealth by shocks, in the absence of alternative coping mechanisms such as credit or transfers. Similarly, the strong influence of diaspora networks and their collapse when local connections are removed reflect the lack of weak ties and heterogeneity in the network topology. While the model exhibits non-linear dynamics, the location and intensity of tipping points depend on the functional-form assumptions and parameter values, particularly those governing memory and risk perception. Finally, the absence of spatial structure, destination preferences, and geographic constraints implies that the results should be interpreted as illustrating underlying mechanisms and potential dynamics, rather than providing quantitative predictions of real-world migration patterns. 
}

% Future work
{
These limitations provide a natural foundation for future work. First, the representation of social networks could be extended to model diaspora connections more explicitly and to capture their co-evolution with migration dynamics. Second, the model could incorporate social and institutional factors that shape agents' capabilities, such as access to credit, policy constraints, or support mechanisms. Third, introducing spatial autocorrelation and geographical constraints would allow for a more realistic representation of the environment. Finally, the model could be calibrated to empirical data and used to evaluate policy interventions, such as cash transfers or information-sharing, thereby enhancing its empirical relevance.  
}

\section*{Acknowledgements}

We are grateful to Prof. Samuel Fankhauser for his guidance and support throughout the development of this article. We would also like to thank Andrea Vismara for helpful comments on an earlier discussion of agency and migration, and Dr. Joanne Burgess Barbier for her careful reading of the manuscript and insightful comments.

\clearpage
\appendix

\section{Appendix}
\subsection{Overview, Design concepts, and Details (ODD) of the Agent-Based Model}
\subsubsection{Time, space, and population}
Time is discrete: $t = 0,1,2,\dots,T$ with step size $\Delta t>0$ and corresponding real time
\[
\tau_t = t\,\Delta t.
\]
The environment is a square grid of cells
\[
\mathcal{G} = \{1,\dots,G\}\times\{1,\dots,G\}, \qquad |\mathcal{G}| = G^2.
\]
There are $N$ agents indexed by $i\in\{1,\dots,N\}$. Each agent occupies a cell
\[
x_i(t) \in \mathcal{G}.
\]

\subsubsection{Cell-level shock process}
Each cell $c\in\mathcal{G}$ is endowed with cell-specific parameters $(\lambda_{0,c}, g_c)$ drawn at initialization:
\[
\lambda_{0,c} \sim \mathcal{U}(\underline{\lambda}_0,\overline{\lambda}_0),
\qquad
g_c \sim \mathcal{U}(\underline{g},\overline{g}).
\]

Define the (nonnegative) shock intensity function in cell $c$ at real time $\tau$:
\[
\lambda_c(\tau) =
\begin{cases}
\max\{0,\lambda_{0,c}\}, & \text{(constant mode)}\\[4pt]
\max\{0,\lambda_{0,c}+\alpha_c\,\tau\}, & \text{(linear mode)}.
\end{cases}
\]

At each discrete step $t$, the number of shock arrivals in cell $c$ is drawn as:
\[
K_c(t) \sim \text{Pois }\!\big(\lambda_c(\tau_t)\big).
\]
The model also stores the cumulative shock count per cell:
\[
E_c(t) = \sum_{s=0}^{t} K_c(s).
\]
Agents observe the shock count in their current location, i.e.
\[
n_i(t)=K_{x_i(t)}(t).
\]
\subsubsection{Agent state variables}
Each agent $i$ has the following state variables at time $t$:
\begin{itemize}
\item perceived risk: $P_i(t)\in\mathbb{R}_{\ge 0}$
\item desire to migrate: $D_i(t)\in\mathbb{R}_{\ge 0}$
\item wealth: $W_i(t)\in\mathbb{R}_{\ge 0}$
\item social capital: $S_i(t)\in\mathbb{R}_{\ge 0}$
\item capability: $C_i(t)\in\mathbb{R}$
\end{itemize}

Each agent is also assigned fixed idiosyncratic parameters (drawn at the initialisation step):
\[
\gamma_i \sim \mathrm{Lognormal}(\mu_{\gamma}, \sigma_{\gamma}),\qquad
\eta_i \sim \mathrm{Lognormal}(\mu_{\eta},\sigma_{\eta}),
\]
where $\gamma_i>0$ is the risk-memory decay (``forgetting'') and $\eta_i\ge 0$ is the agent-specific sensitivity to shocks.

In addition, each agent draws: 
\[
\kappa_i \sim \mathcal{U}(\underline{\kappa}, \overline{\kappa}), 
\qquad
\alpha_{W,i} \sim \mathcal{U}(\underline(\alpha)_W, \overline{\alpha}_W), 
\qquad
\delta_{W,i} \sim \mathcal{U}(\underline{\delta}_W, \overline{\delta}_W), 
\]
where $\kappa_i$ governs post-migration risk updating, $\alpha_{W,i}$ controls wealth persistence, and $\delta_{W,i}$ determines proportional wealth loss under shock exposure. 

Initial wealth is drawn from a lognormal distribution with fixed mean level $\bar W$ and dispersion parameter $\sigma_W$: 

\[
W_i(0) \sim \mathrm{Lognormal}(\mu_W, \sigma_W),
\qquad
\mu_W = \log(\bar W) - \frac{1}{2} \sigma_W^2,
\]
where $W_{i,0} = W_i(0)$ denote the initial wealth. 

\subsubsection{Initial conditions}

Agents are initially assigned to cells uniformly at random. Given the baseline shock intensity in their initial location, agents are initialised with 
\[
P_i(0) = \max \left\{0, \frac{\eta_i \lambda_{0, x_i(0)}}{\gamma_i} \right\}, 
\qquad
D_i(0) = P_i(0),
\qquad
S_i(0)
\]

Capability is initialised as:
\[
C_i(0) = \psi_W W_i(0) + \psi_S S_i(0)
\]

The initialisation approximates the local steady-state risk level implied by the baseline arrival intensity. 

\subsubsection{Risk perception dynamics}
Perceived risk evolves according to the following differential equation. Thus, recent shock exposure raises perceived risk, while past risk decays at an agent-specific rate $\gamma_i$.

\begin{equation}
\frac{d P_i}{d\tau} = -\gamma_iP_i(\tau) + \eta_i \xi_i(\tau),
\label{eq:perceptionappendix}
\end{equation}

While $\xi_i(\tau)$ is the instantaneous shock exposure signal in continuous time, $n_i(t)$ is the number of shocks in a discrete time step. Continuous-time equations are implemented numerically using discrete-time updates with step size $\Delta t$. 

\subsubsection{Desire dynamics}
Desire follows a linear relaxation toward perceived risk:
\[
\frac{dD_i}{d\tau} = r(P_i(\tau) - D_i(\tau))
\]
with base growth rate $r>0$.

\subsubsection{Wealth dynamics with heterogeneous vulnerability and resilience}

Let $W_{i,0}$ denote agent $i$'s baseline (pre-shock) wealth, equal to initial wealth.
Agents differ along two dimensions:
\begin{itemize}
\item \emph{vulnerability to shocks}, $\delta_{W,i}\in(0,1)$,
\item \emph{wealth persistence} (inverse resilience), $\alpha_{W,i}\in(0,1)$.
\end{itemize}
Both parameters are drawn independently at initialisation.

At time $t$, agent $i$ experiences $n_i(t)$ shocks in its current cell. Wealth evolves as:
\[
W_i(t+1)=
\begin{cases}
(1-\delta_{W,i})\,W_i(t), & \text{if } n_i(t) > 0,\\[6pt]
\alpha_{W,i}\,W_i(t) + \big(1-\alpha_{W,i}\big)\,W_{i,0}, & \text{if } n_i(t)=0.
\end{cases}
\]

The first case represents a proportional wealth loss upon exposure to at least one shock, with heterogeneity in vulnerability across agents. The second case captures partial recovery toward baseline wealth in periods without shocks. Lower values of $\alpha_{W,i}$ correspond to faster recovery (higher resilience). Wealth is constrained to remain non-negative.

\subsubsection{Social capital from locals and diaspora}

After migration decisions (see below), social capital is updated using counts of locals and diaspora.

Let
\[
\mathcal{I}_c(t) = \{i: x_i(t)=c\}
\]
be the set of agents located in cell $c$ after moves at step $t$. For an agent $i$ currently in cell $x_i(t)$, define:
\[
L_i(t) = |\mathcal{I}_{x_i(t)}(t)| \quad \text{(locals count)}.
\]

Let $x_i(t-1)$ be the agent's previous location (before moving in step $t$). Define the set of realised origin-destination pairs among movers:
\[
\mathcal{M}(t) = \{(o,d)\in\mathcal{G}\times\mathcal{G}:\exists i \text{ with } x_i(t-1)=o,\ x_i(t)=d,\ d\neq o\}.
\]
For each origin $o$, define the origin-to-diaspora size (as coded: sum of destination population sizes over distinct destinations reached from $o$ that step):
\[
\text{DiasporaSize}(o,t) = \sum_{(o,d)\in\mathcal{M}(t)} |\mathcal{I}_d(t)|.
\]
Then for agent $i$ with origin $o_i(t)=x_i(t-1)$,
\[
\mathrm{Di}_i(t) = \text{DiasporaSize}(o_i(t),t).
\]

Convert these counts into shares of the total population $N$:
\[
\ell_i(t)=\frac{L_i(t)}{N},\qquad
\delta_i(t)=\frac{\mathrm{Di}_i(t)}{N}.
\]

Social capital updates via:
\[
\frac{d S_i}{d\tau} = -\gamma_S S_i(\tau) + \beta_L \ell_i(\tau) + \beta_D \delta_i(\tau)
\]
where $\gamma_S\ge 0$ is the social-capital decay and $\beta_L,\beta_D\ge 0$ are the weights of locals and the flows of the diaspora.

\subsubsection{Capability dynamics}
Agents maintain a capability stock $C_i(t)$. Capability evolves using a weighted flow update based on wealth and social-capital derivatives:
\[
\frac{d C_i}{d\tau} = \psi_W \frac{d W_i}{d\tau} + \psi_S \frac{dS_i}{d\tau},
\]
where $\psi_W,\psi_S\ge 0$ are flow weights ( with $\psi_W+\psi_S=1$).

In the implementation, capability is updated before migration decisions using current wealth and the most recently available social capital, and is recomputed only in the step following social capital's update. 

\subsubsection{Migration decision rule}
At each time step, agent $i$ is classified into one of four aspiration-capability categories based on thresholds $\phi$ (desire) and $\theta$ (capability):
\[
\begin{array}{ll}
\text{Migrant state:} & D_i(t)\ge \phi \ \land\ C_i(t)\ge \theta,\\
\text{Trapped:} & D_i(t)\ge \phi \ \land\ C_i(t)< \theta,\\
\text{Voluntary non-migrant:} & D_i(t)< \phi \ \land\ C_i(t)\ge \theta,\\
\text{Acquiescent:} & D_i(t)< \phi \ \land\ C_i(t)< \theta.
\end{array}
\]

The behavioural migration decision is:
\[
m_i(t) =
\mathbf{1}\{D_i(t)\ge\phi \ \land\ C_i(t)\ge\theta\}.
\]
If $m_i(t)=1$, agent $i$ relocates by drawing a new cell uniformly from all cells other than the current one:
\[
x_i(t+1)\sim \mathcal{U}\bigl(\mathcal G \setminus\{x_i(t)\}\bigr).
\]
Otherwise,
\[
x_i(t+1)=x_i(t).
\]

After migration, perceived risk is partially adjusted toward the expected risk level of the destination cell. Let $\lambda_{x_i(t+1)}(\tau_t)$ denote the destination intensity at time $\tau$, and define
\[
P_i^{*}(t)=\frac{\eta_i\,\lambda_{x_i(t+1)}(\tau_t)\Delta t}{1-e^{-\gamma_i\Delta t}}.
\]
Then the post-move risk adjustment is
\[
P_i(t+1)= \kappa_i P_i(t+1)+(1-\kappa_i)P_i^{*}(t),
\]
followed by
\[
D_i(t+1) = P_i(t+1).
\]

\subsubsection{Update order (per step)}
For each step $t=0,\dots,T-1$:
\begin{enumerate}
\item Draw shock counts $K_c(t)$ for all cells $c$.
\item For each agent, $i$ at current location $x_i(t)$:
  update perceived risk $P_i$, desire $D_i$, and wealth $W_i$ using the shocks in the current cell $n_i(t)=K_{x_i(t)}(t)$.
\item Update capability $C_i$ from current wealth and social capital.
\item Apply migration rule to update locations.
\item Compute locals/diaspora counts after moves and update social capital $S_i$.
\item Record aggregate outcomes and agent trajectories.
\end{enumerate}

\subsubsection{Parameters}

\begin{table}[H]
\centering
\small
\caption{Baseline model parameters}
\label{tab:baseline_parameters}

\begin{tabularx}{\textwidth}{l l l X}
\toprule
\textbf{Category} & \textbf{Parameter} & \textbf{Value} & \textbf{Description} \\
\midrule

Environment 
& $G$ 
& 10 
& Grid size of the spatial environment \\

Population 
& $N$ 
& 100 
& Number of agents \\

Simulation horizon 
& $T$ 
& $100 \times 52$ 
& Number of weekly simulation steps \\

Time step 
& $\Delta t$ 
& $1/52$ 
& One week expressed in years \\

Capability weights 
& $(w_1,w_2)$ 
& $(0.6,0.4)$ 
& Weights used in capability elasticity \\

Capability composition 
& $(\psi_W,\psi_S)$ 
& $(0.6,0.4)$ 
& Weights on wealth and social capital \\

Social capital decay 
& $\gamma_S$ 
& 0.01 
& Decay rate of social capital \\

Risk memory 
& $(\mu_\gamma,\sigma_\gamma)$ 
& $(-2.5,0.8)$ 
& Lognormal risk-memory distribution parameters \\

Network formation 
& $(\beta_L,\beta_D)$ 
& $(0.4,0.6)$ 
& Local and diaspora network weights \\

Wealth recovery 
& $\alpha_W$ 
& $(0.7,0.95)$ 
& Wealth recovery range \\

Shock losses 
& $\delta_W$ 
& $(0.05,0.4)$ 
& Fractional wealth loss from shocks \\

Desire adjustment 
& $r$ 
& 0.1 
& Speed of desire adaptation \\

Migration sensitivity 
& $\kappa$ 
& $(0.3,0.9)$ 
& Desire sensitivity range \\

Migration thresholds 
& $(\phi,\theta)$ 
& $(0.5,0.5)$ 
& Desire and capability thresholds for migration \\

Shock frequency 
& $\lambda_0$ 
& $(0.01,0.1)$ 
& Baseline shock arrival rate \\

Shock growth 
& $g$ 
& $(0.00003,0.001)$ 
& Growth rate of climate shocks \\

Shock sensitivity 
& $(\mu_\eta,\sigma_\eta)$ 
& $(-0.8,0.8)$ 
& Lognormal shock-sensitivity parameters \\

Initial wealth 
& $(\mu_W,\sigma_W)$ 
& $(-0.7,0.2)$ 
& Initial lognormal wealth distribution \\

\bottomrule
\end{tabularx}
\end{table}

\subsection{Results}
\subsubsection{State comparison of mobility states over time}
\begin{figure}[H] 
\centering \includegraphics[width=\textwidth]{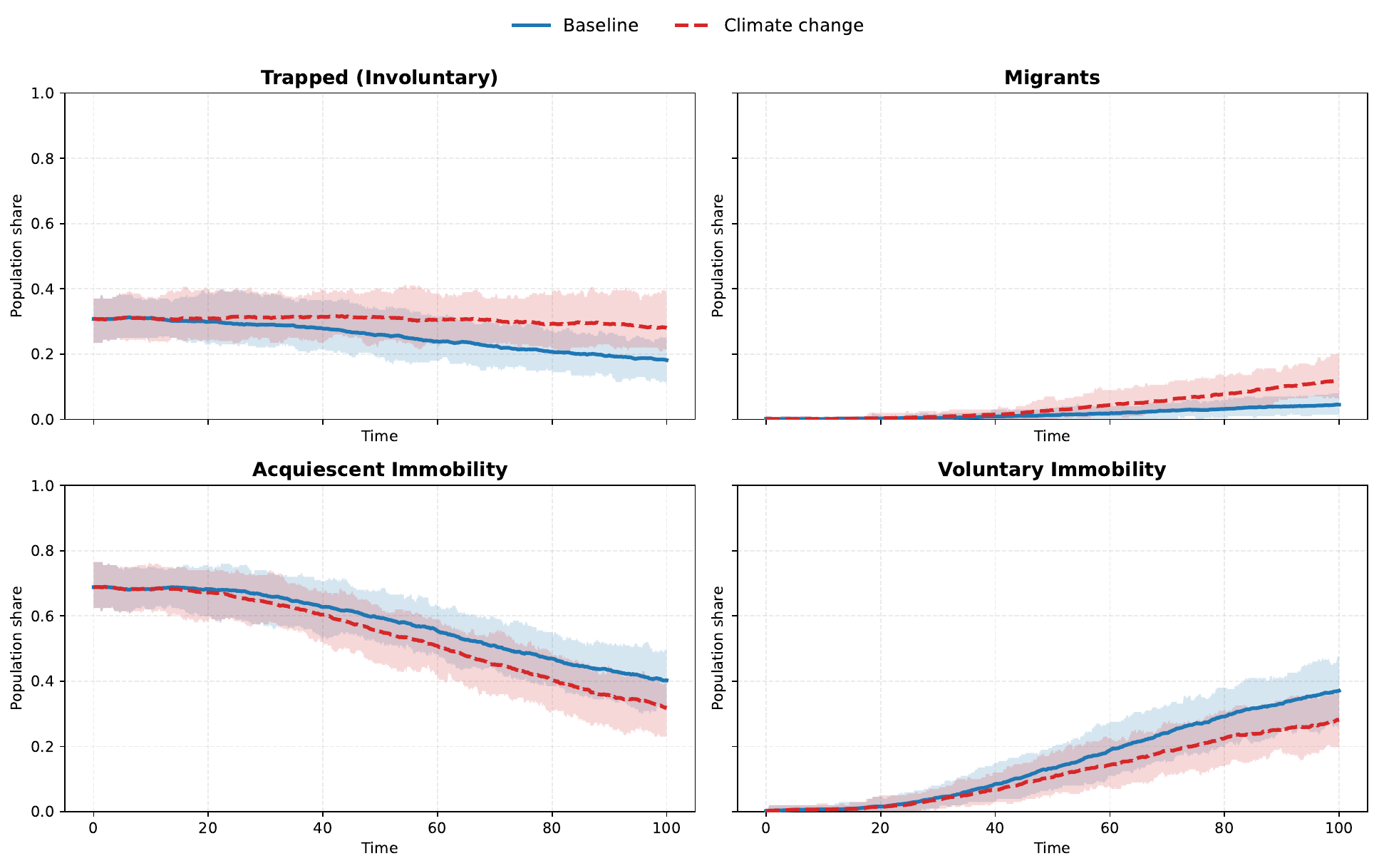} \vspace{0.3cm}  
\caption{State comparison of mobility states over time} 
\label{fig:wealthshocksandrecovery} 
\end{figure}

\begin{figure}[H] 
\centering 
\includegraphics[width=15cm,trim=0 0cm 0 0cm,clip]{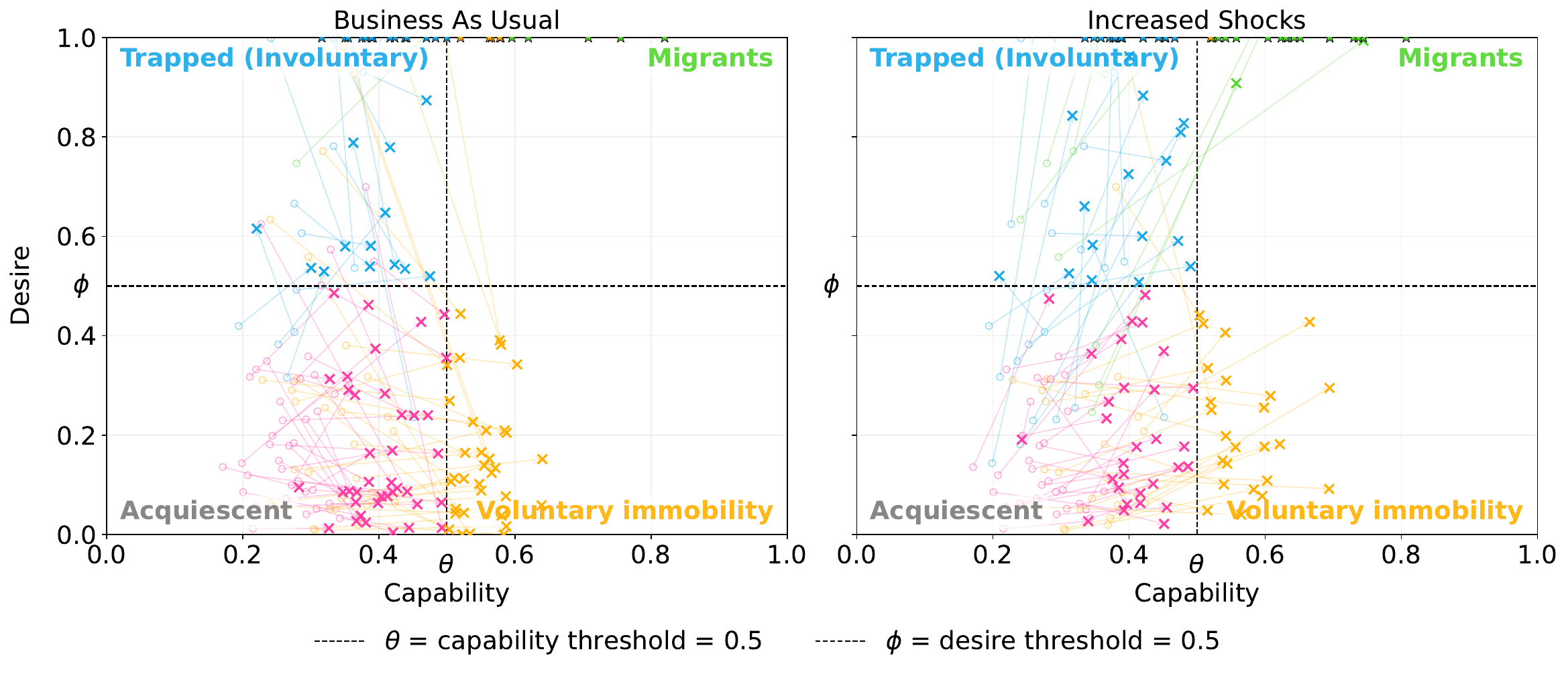} 
\caption{Agent Start-End Displacements in Desire-Capability Space for the Business as Usual scenario (left) and shocked-climate scenario (right)} 
\label{fig:trajectories} 
\end{figure}

\subsubsection{Changing wealth shock (\texorpdfstring{$\delta_{w,i}$}{delta	extunderscore{w,i}}) and wealth recovery (\texorpdfstring{$\alpha_{w,i}$}{alpha	extunderscore{w,i}})}

\begin{table}[H]
\centering
\small
\caption{Economic regimes defined by wealth shocks ($\delta_{w,i}$) and recovery rates ($\alpha_{w,i}$)}
\label{tab:regimes}

\begin{tabularx}{\textwidth}{l l X}
\hline
\textbf{Regime} & \textbf{Configuration} & \textbf{Interpretation} \\
\hline

Resilient
& Low $\delta_{w,i}$, high $\alpha_{w,i}$
& Fast recovery \\

Poverty trap
& High $\delta_{w,i}$, low $\alpha_{w,i}$
& Persistent capability loss \\

Adaptive but volatile
& High $\delta_{w,i}$, high $\alpha_{w,i}$
& Frequent shocks with recovery \\

Stable but slow
& Low $\delta_{w,i}$, low $\alpha_{w,i}$
& Slow economic adjustment \\

\hline
\end{tabularx}
\end{table}

\begin{figure}[H] 
\centering \includegraphics[width=\textwidth]{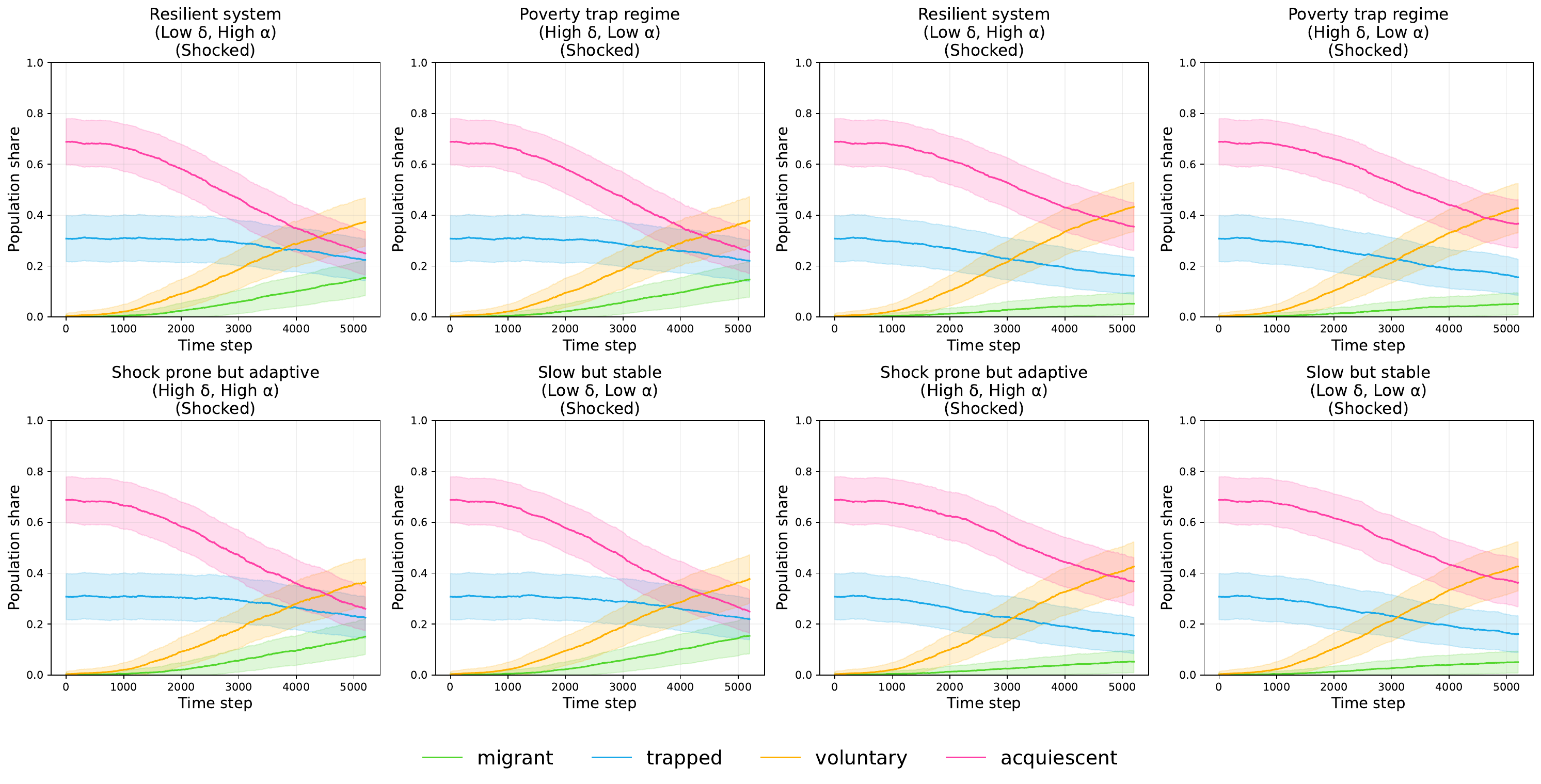}  \caption{Mobility outcomes by shock intensity and recovery capacity} \label{fig:regime_comparison} \end{figure} 

\subsubsection{The effect of desire adjustment on migration states}

\begin{figure}[H] 
\centering 
\includegraphics[width=\textwidth]{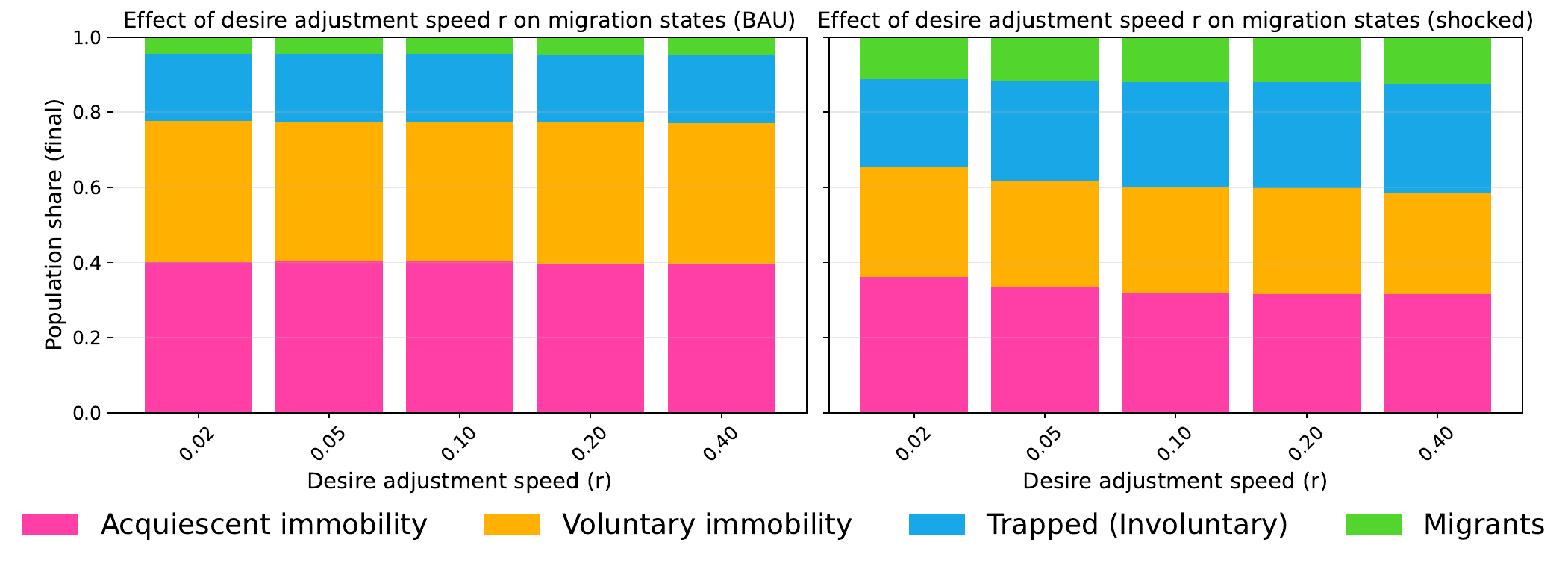} 
\caption{} 
\label{fig:rsensitivity} \end{figure} 

\subsubsection{Intensity of arrival of climate events}

\begin{figure}[H] 
\centering 
\includegraphics[width=13cm]{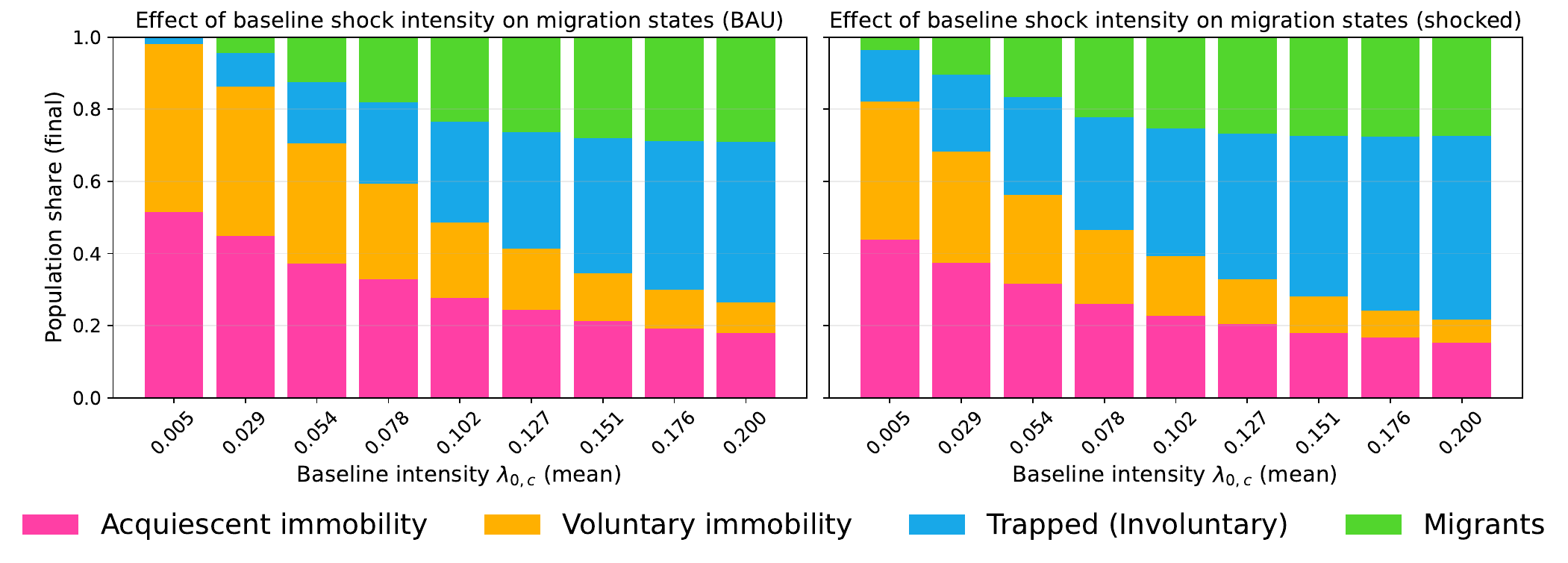} 
% \vspace{0.3cm}  
\caption{} 
\label{fig:lambda_sensitivity} 
\end{figure} 

\subsubsection{Effects on risk perception: mean memory decay \texorpdfstring{$\gamma_i$}{gamma	extsubscript{i}} and mean shock sensitivity \texorpdfstring{$\eta_i$}{eta	extsubscript{i}}}

\begin{figure}[H] 
\centering 
\includegraphics[width=13cm]{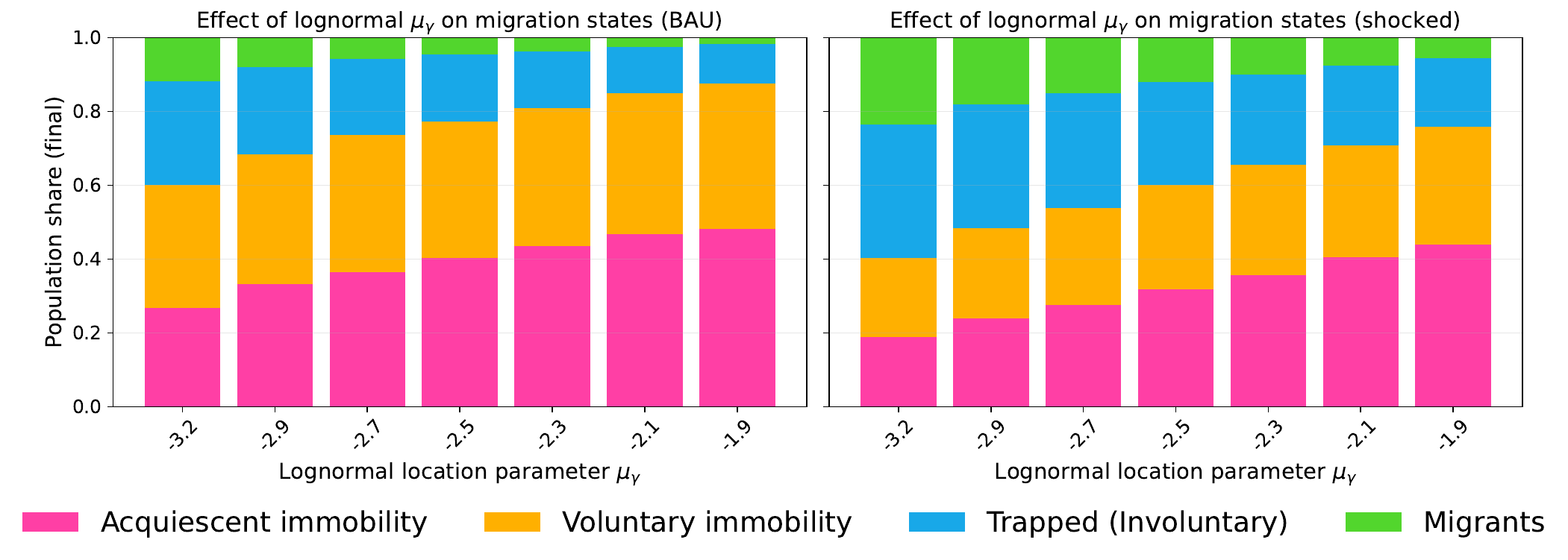} 
% \vspace{0.3cm}  
\caption{} 
\label{fig:gamma_sensitivity} 
\end{figure} 

\begin{figure}[H] 
\centering 
\includegraphics[width=13cm]{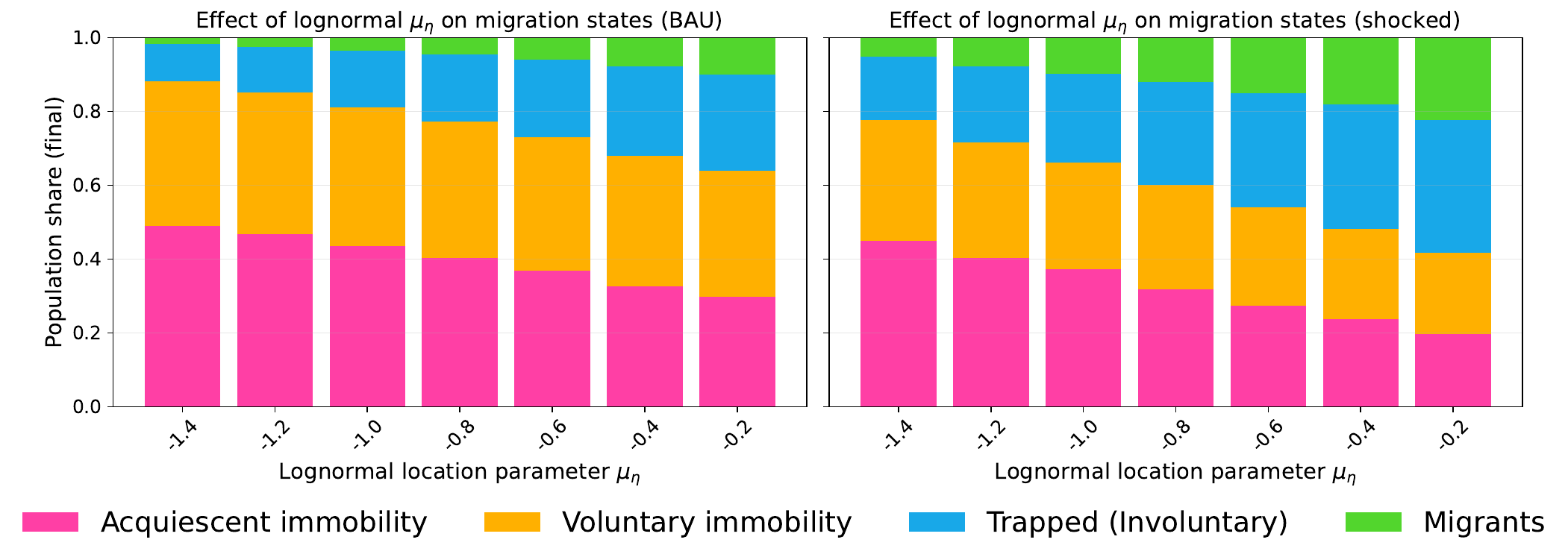} 
% \vspace{0.3cm}  
\caption{} 
\label{fig:eta_sensitivity} 
\end{figure}

\subsubsection{Climate Change, Migration and the Diaspora}

% Psi (entire social network - local and diaspora)

{
When analyzing the effects of the Network on shaping migration capability, we find that it only produces modest changes in mobility outcomes. As $\psi_S$ rises, acquiescent immobility declines slightly, with corresponding increases in voluntary immobilty, and to a lesser extent, trapped and migrant populations. This effect is stronger under the shocked climate scenario, where the networks partially buffer capability losses. Overall, these results suggest that expanding the influence of social capital relaxes mobility constraints at the margin, but does not fundamentally alter the distribution of mobility outcomes. 

\begin{figure}[H] 
\centering 
\includegraphics[width=13cm]{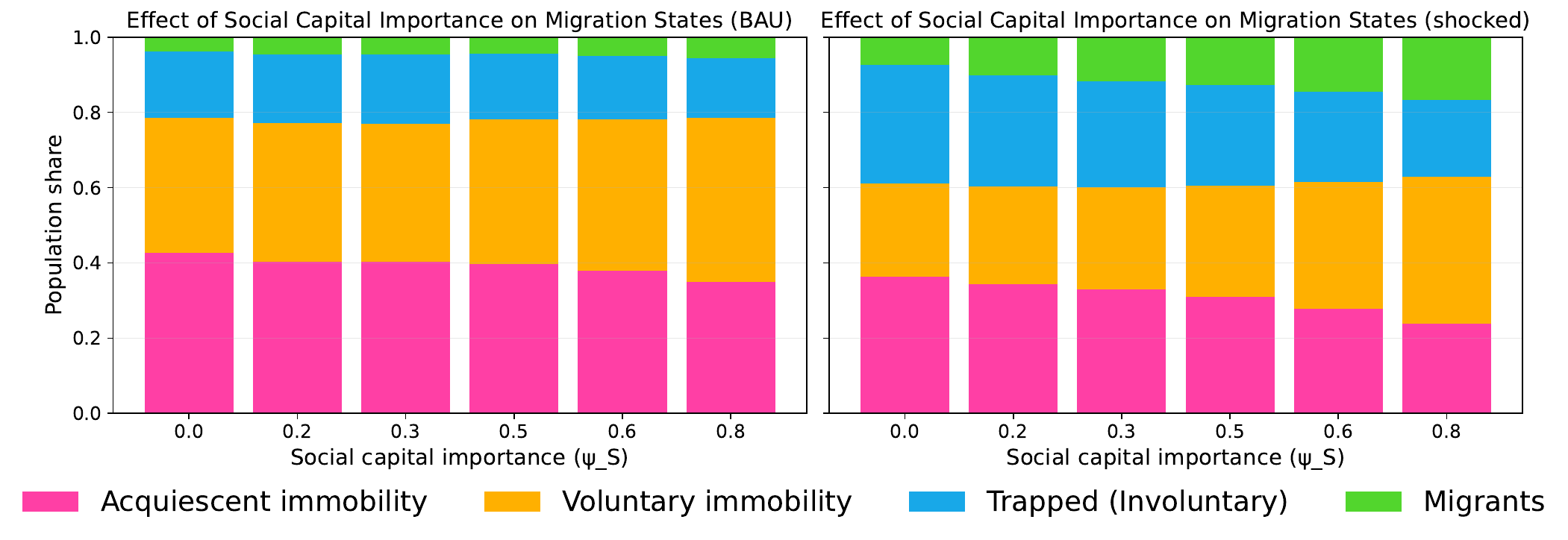} 
% \vspace{0.3cm}  
\caption{} 
\label{fig:psi_sensitivity} 
\end{figure} 
}

% \beta_D (network composition - just the diaspora)

{
We then look at the effects of diaspora specifically, to see if changes in the network composition would generate pronounced shifts. As $\beta_D$ increases, voluntary immobility declines sharply, while both acquiescent and trapped populations expand substantially, alongside a more modest rise in migration. This pattern is particularly strong under climate shocks. The results that indicate that diaspora-based networks do not uniformly relax mobility constraints; instead, they redistribute access to migration opportunities. While some agents benefit from their diaspora connections and are able to migrate, others,  lacking social ties in a different cell, become relatively more constrained, leading to increased polarization between mobile and immobile populations. 

At extreme values where social capital is entirely derived from diaspora networks, overall capability declines for a large share of the population. This reflects the fact that diaspora connections are not uniformly accessible: while they can strongly enhance migration opportunities for connected individuals, they provide little support to others. In the absence of broad-based local networks, this leads to a concentration of agents in acquiescent and trapped states, highlighting the importance of network accessibility rather than network presence alone.

\begin{figure}[H] 
\centering 
\includegraphics[width=13cm]{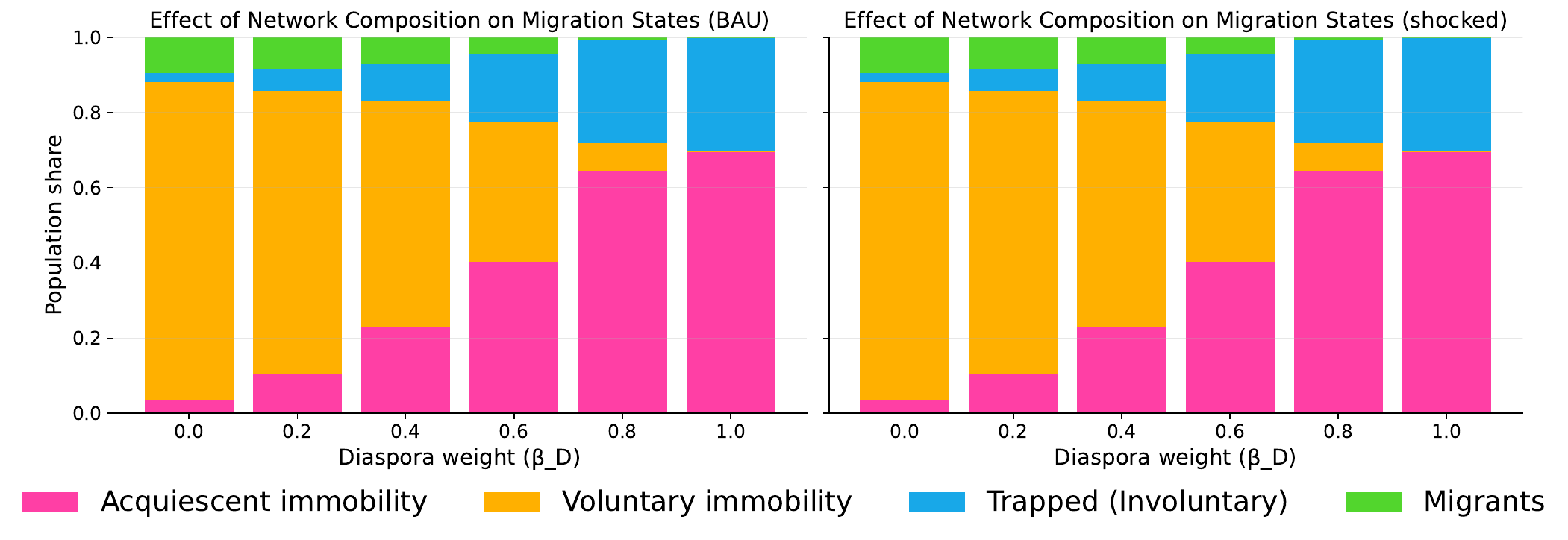} 
% \vspace{0.3cm}  
\caption{} 
\label{fig:beta_sensitivity} 
\end{figure} 
}

So, while stronger networks modestly ease mobility constraints, the diaspora part of the network fundamentally reshape who is able to move by unevenly distributing access to migration.

\subsubsection{Stress testing results}

\begin{table}[htbp]
\centering
\caption{Baseline and climate scenario outcomes across 50 stochastic seeds.}
\label{tab:scenario_stress_summary_full}
\begin{tabular}{lcc}
\hline
Scenario & Cumulative shocks & Cumulative moves \\
\hline
Baseline & 545.26 $\pm$ 34.03 [480, 625] & 8762.98 $\pm$ 4541.90 [694, 19246] \\
Climate  & 559.10 $\pm$ 32.56 [502, 631] & 9867.68 $\pm$ 5117.19 [1165, 20533] \\
\hline
\end{tabular}
\end{table}

\begin{table}[p]
\centering
\footnotesize
\caption{Stress-test outcomes across alternative parameter regimes. Means are reported in the first row and standard deviations in parentheses below.}
\label{tab:stress_tests}

\renewcommand{\arraystretch}{1.1}

\begin{tabular}{lcccccc}
\hline
Scenario & Migrant & Trapped & Voluntary & Acquiescent & Cum.\ moves & Cum.\ shocks \\
\hline

Baseline
& 0.044 & 0.189 & 0.367 & 0.400 & 9315.9 & 563.3 \\
& \scriptsize(0.022)
& \scriptsize(0.039)
& \scriptsize(0.069)
& \scriptsize(0.060)
& \scriptsize(5422.4)
& \scriptsize(34.2)
\\[0.15cm]

Combined stress
& 0.000 & 0.629 & 0.000 & 0.371 & 0.0 & 2118.3 \\
& \scriptsize(0.000)
& \scriptsize(0.050)
& \scriptsize(0.000)
& \scriptsize(0.050)
& \scriptsize(0.0)
& \scriptsize(90.6)
\\[0.15cm]

Fast memory decay
& 0.026 & 0.111 & 0.381 & 0.482 & 5205.5 & 568.3 \\
& \scriptsize(0.017)
& \scriptsize(0.033)
& \scriptsize(0.051)
& \scriptsize(0.052)
& \scriptsize(4154.4)
& \scriptsize(30.4)
\\[0.15cm]

High shock pressure
& 0.184 & 0.467 & 0.145 & 0.203 & 31131.8 & 1474.1 \\
& \scriptsize(0.043)
& \scriptsize(0.064)
& \scriptsize(0.050)
& \scriptsize(0.033)
& \scriptsize(9616.1)
& \scriptsize(72.3)
\\[0.15cm]

High shock sensitivity
& 0.113 & 0.272 & 0.329 & 0.286 & 23182.3 & 567.3 \\
& \scriptsize(0.030)
& \scriptsize(0.049)
& \scriptsize(0.063)
& \scriptsize(0.052)
& \scriptsize(8342.5)
& \scriptsize(34.3)
\\[0.15cm]

Socially networked
& 0.004 & 0.298 & 0.023 & 0.675 & 1255.1 & 565.9 \\
& \scriptsize(0.006)
& \scriptsize(0.042)
& \scriptsize(0.016)
& \scriptsize(0.039)
& \scriptsize(2014.0)
& \scriptsize(36.0)
\\[0.15cm]

Tight thresholds
& 0.002 & 0.223 & 0.039 & 0.736 & 221.7 & 567.3 \\
& \scriptsize(0.006)
& \scriptsize(0.039)
& \scriptsize(0.024)
& \scriptsize(0.033)
& \scriptsize(636.6)
& \scriptsize(37.0)
\\

\hline
\end{tabular}

\begin{flushleft}
\footnotesize
\textit{Note:} Final shares refer to the proportion of agents in each state at the final time step. Cumulative moves denote the total number of migration events over the full simulation horizon. Cumulative shocks denote the total number of shock arrivals across all cells and time steps.
\end{flushleft}

\end{table}

The stress test show that the model responds coherently to extreme parameter configurations and that different mechanisms generate distinct forms of mobility and immobility. The high shock-pressure scenario produces the strongest mobility response, with both cumulative moves and the final migrant share increasing substantially relative to baseline. However, this increase in movement is accompanied by a large rise in trapped immobility and a sharp decline in voluntary immobility. This indicates that intensified environmental exposure does not simply produce more migration; it also pushes a larger share of agents into situations where they want to move but lack sufficient capability to do so. 

The high shock-sensitivity scenario produces a similar but less extreme pattern. When agents respond more strongly to shocks, migration events increase and the trapped population expands, even though the underlying number of shocks remains close to the baseline. This suggests that behavioural responses to risk can amplify the consequences of a given level of environmental exposure. In contrast, scenarios that restrict capability or weaken mobility channels suppress realised migration. Tight thresholds reduce migration almost completely and shift the population toward acquiescent and trapped states. Similarly, the socially networked scenario produces very low migration and large acquiescent population, suggesting that when social support is insufficient to overcome capability constraints, immobility becomes dominant outcome. 

The combined-stress scenario represents the most severe mobility trap. Despite extremely high cumulative shock exposure, no realised migration occurs. Instead, the population is divided between trapped and acquiescent immobility. This outcome illustrates the central mechanism of the model: shocks can increase migration pressure while simultaneously eroding  the resources required to migrate. Under sufficiently adverse conditions, environmental stress therefore produces involuntary immobility rather than mass movement. Overall, the stress tests demonstrate that climate-related mobility is shaped by the balance between exposure, behavioural response, and capability constraints, with extreme stress generating both migration and trapping depending on whether agents retain the means to act on their desire to move. 

\subsubsection{Robustness table on \texorpdfstring{$\Delta t$}{Delta t} and population size}

\begin{figure}[H] 
\centering 
\includegraphics[width=16cm]{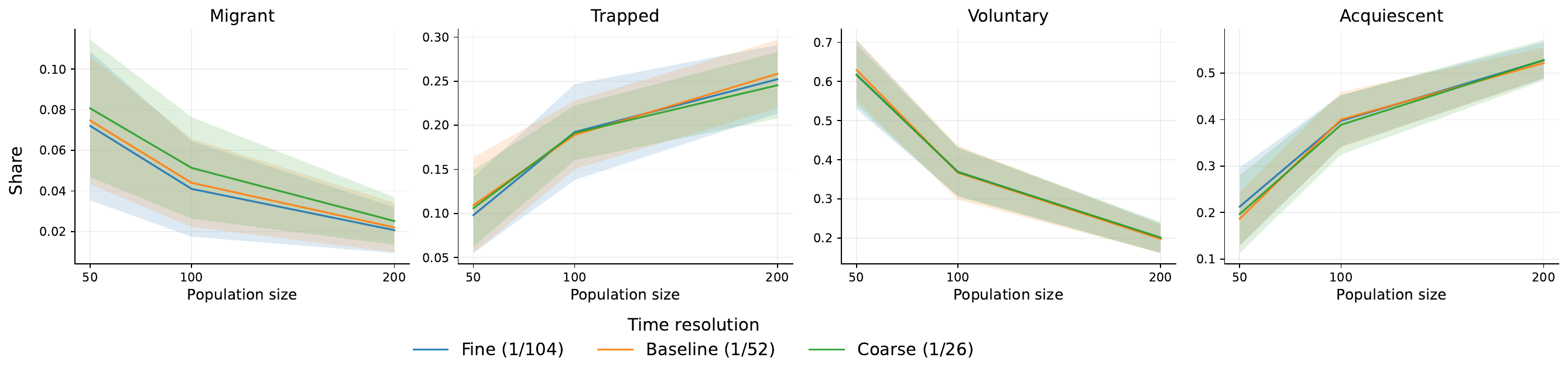} 
% \vspace{0.3cm}  
\caption{Robustness of migration outcomes to time discretisation and population size. The lines show mean values across stochastic simulations, with shaded bands indicating one standard deviation across seeds.} 
\label{fig:robustness} 
\end{figure}

In figure \ref{fig:robustness} we present the results of the robustness checks for time discretisation and population size. Results are averaged across stochastic runs, with shaded bands indicating one standard deviation. The model exhibits clear scaling behaviour with respect to population size: as the number of agents increases, the final share of migrants declines while the share of trapped individuals rises, accompanied by a reduction in voluntary mobility. These trends are smooth and monotonic across all specifications. In contrast, variation in the time step has negligible impact on outcomes. The near overlap of trajectories across different values of dt, together with overlapping uncertainty bands, indicates that the results are not driven by numerical discretisation. Overall, the figure demonstrates that the model's core dynamics are robust to implementation choices while exhibiting meaningful and systematic scaling effects. This suggests that mobility constraints emerge end intensify with system size.

\subsubsection{Sensitivity Analysis}

We first simulated the model under baseline and climate-change scenarios across multiple stochastic seeds. We then used Latin hypercube sampling to explore behaviour across a broad multidimensional parameter space and screened input-output relationships using Spearman rank correlations \citep{mckay1979latin, spearman1904association}. Finally, we performed a Sobol global sensitivity analysis on a subset of key parameters, averaging outcomes across repeated stochastic realizations at each design point to reduce Monte Carlo noise \citep{sobol2001global, herman2017salib, saltelli2010variance} . Together, these approaches provide complementary perspectives on model sensitivity: the Latin hypercube analysis offers a broad screening of monotonic associations across a wide parameter set, while the Sobol analysis provides a more formal ranking of parameter importance within a smaller subset of key inputs.

\paragraph{Latin Hypercube Screening Analysis and Spearman correlation measure}

\begin{table}[htbp]
\centering
\footnotesize
\setlength{\tabcolsep}{4pt}

\caption{Top three parameter associations from the Latin hypercube screening analysis.}
\label{tab:lhs_top3}

\begin{tabular}{llll}
\toprule
Outcome & Rank 1 & Rank 2 & Rank 3 \\
\midrule

Cumulative moves 
& $\theta$ ($\rho=-0.55$) 
& $\gamma_S$ ($\rho=-0.39$) 
& $\beta_D$ ($\rho=-0.33$) \\

Final migrant share 
& $\theta$ ($\rho=-0.48$) 
& $\gamma_S$ ($\rho=-0.40$) 
& $\mu_{\gamma}$ ($\rho=-0.32$) \\

Final trapped share 
& $\mu_{\gamma}$ ($\rho=-0.44$) 
& $\theta$ ($\rho=0.37$) 
& $\beta_D$ ($\rho=0.31$) \\

Final voluntary share 
& $\theta$ ($\rho=-0.68$) 
& $\gamma_S$ ($\rho=-0.57$) 
& $\beta_D$ ($\rho=-0.41$) \\

Final acquiescent share 
& $\theta$ ($\rho=0.59$) 
& $\gamma_S$ ($\rho=0.53$) 
& $\beta_D$ ($\rho=0.30$) \\

\bottomrule
\end{tabular}

\vspace{0.2cm}

\begin{flushleft}
\footnotesize
\textit{Note:} Results are based on a Latin hypercube exploration of the multidimensional parameter space. Reported coefficients are Spearman rank correlations. All reported correlations are statistically significant at the 1\% level.
\end{flushleft}

\end{table}

The Latin hypercube screening \citep{mckay1979latin} indicates that the capability threshold ($\theta$) and social-capital decay ($\gamma_S$) are the parameters most consistently associated with model outcomes across the sampled parameter space. Both parameters are negatively associated with cumulative moves, final migrant share, and final voluntary share, while being positively associated with final acquiescent share, consistent with stronger capability constraints reducing realized mobility. The diaspora effect ($\beta_D$) also appears among the leading correlates for several outcomes, suggesting that stronger network support facilitates migration. The final trapped share differs from the other outcomes in being most strongly associated with mean memory decay ($\mu_{\gamma}$), pointing to a particular role for the persistence of perceived risk in generating trapped populations. As these are rank correlations, the table should be interpreted as a screening exercise that identifies broad monotonic associations rather than causal effect sizes or variance shares.

\paragraph{Sobol Sensitivity Analysis}

\begin{table}[htbp]
\centering
\caption{Sobol total-order sensitivity rankings for key model outcomes. Parameters are ordered by total-order index ($ST$), which captures both direct and interaction effects. Given estimator instability in first-order indices, results are interpreted primarily in terms of relative ranking rather than exact variance shares.}
\label{tab:sobol_rankings}

\begin{tabular}{lllc}
\hline
Outcome & Rank & Parameter & $ST$ [95\% CI] \\
\hline

Cumulative moves & 1 & $\theta$       & 1.321 [0.645, 1.997] \\
                 & 2 & $\mu_{\gamma}$ & 1.093 [0.569, 1.618] \\
                 & 3 & $\lambda_0$    & 1.091 [0.611, 1.570] \\
                 & 4 & $\beta_D$      & 1.071 [0.343, 1.798] \\
                 & 5 & $\gamma_S$     & 1.016 [0.558, 1.474] \\
                 & 6 & $\mu_{\eta}$   & 0.985 [0.607, 1.362] \\

\hline
Final acquiescent share & 1 & $\lambda_0$    & 1.111 [0.884, 1.339] \\
                        & 2 & $\gamma_S$     & 1.092 [0.856, 1.327] \\
                        & 3 & $\mu_{\gamma}$ & 1.038 [0.812, 1.263] \\
                        & 4 & $\mu_{\eta}$   & 1.013 [0.789, 1.237] \\
                        & 5 & $\theta$       & 1.001 [0.780, 1.221] \\
                        & 6 & $\beta_D$      & 0.928 [0.736, 1.119] \\

\hline
Final migrant share & 1 & $\theta$       & 1.370 [0.793, 1.947] \\
                    & 2 & $\mu_{\gamma}$ & 1.144 [0.712, 1.576] \\
                    & 3 & $\lambda_0$    & 1.093 [0.729, 1.457] \\
                    & 4 & $\mu_{\eta}$   & 1.081 [0.723, 1.439] \\
                    & 5 & $\gamma_S$     & 1.034 [0.621, 1.448] \\
                    & 6 & $\beta_D$      & 0.973 [0.619, 1.328] \\

\hline
Final trapped share & 1 & $\gamma_S$     & 1.182 [0.903, 1.460] \\
                    & 2 & $\mu_{\eta}$   & 1.181 [0.911, 1.450] \\
                    & 3 & $\mu_{\gamma}$ & 1.105 [0.856, 1.353] \\
                    & 4 & $\beta_D$      & 1.091 [0.845, 1.337] \\
                    & 5 & $\lambda_0$    & 1.015 [0.760, 1.270] \\
                    & 6 & $\theta$       & 1.006 [0.781, 1.231] \\

\hline
Final voluntary share & 1 & $\lambda_0$    & 1.088 [0.866, 1.309] \\
                      & 2 & $\mu_{\gamma}$ & 1.054 [0.784, 1.325] \\
                      & 3 & $\gamma_S$     & 1.052 [0.838, 1.266] \\
                      & 4 & $\beta_D$      & 1.001 [0.787, 1.214] \\
                      & 5 & $\mu_{\eta}$   & 0.990 [0.749, 1.232] \\
                      & 6 & $\theta$       & 0.940 [0.712, 1.167] \\

\hline
\end{tabular}
\end{table}

The Sobol sensitivity analysis \citep{sobol2001global, herman2017salib, saltelli2010variance} broadly reinforces the patterns identified by the Latin hypercube screening and Spearman rank correlations \citep{mckay1979latin, spearman1904association}, while providing a more formal measure of total parameter importance that accounts for nonlinearities and interaction effects. For cumulative moves and the final migrant share, the most influential parameters are, the capability threshold ($\theta$), mean memory decay ($\mu_{\gamma}$), and the baseline shock arrival rate ($\lambda_0$). This suggests that aggregate mobility patterns depend not only on migration constraints, but also on the persistence of perceived risk and the overall level of climate exposure experienced by agents. Cumulative moves capture the total number of migration events occurring throughout the simulation, whereas the final migrant share measures the proportion of agents classified as migrants at the final step. The former therefore reflects overall mobility intensity, while the latter captures the long-run composition of the population after mobility processes have unfolded. 

The relative importance of parameters differs across mobility outcomes Final acquiescent and voluntary shares are more strongly influenced by the shock process, with $\lambda_0$, $\mu_{\gamma}$, and $\gamma_S$ consistently ranking among the most important factors. This indicates that long-run immobiilty outcomes depend heavily on how frequently shocks occur and how strongly they shape perceptions and social capital over time. In contrast, the final trapped share is most sensitive to social capital decay ($\gamma_S$), mean shock sensitivity ($\mu_{\eta}$), and mean memory decay ($\mu_{\gamma}$). This pattern is consistent with the model's underlying mechanism: persistent perceptions of risk can increase migration desire, while repeated shocks and weakening social support simultaneously reduce migration capability, increasing the likelihood of becoming trapped. 

Taken together, the Sobol analysis confirms that mobility outcomes emerge from the interaction of climate exposure, risk perception, memory, social capital dynamics, and migration constraints rather than from any single parameter in isolation. The results also suggest that behavioural and social mechanisms are at least as important as environmental forcing in shaping long-run mobility outcomes. 

The Sobol estimates should be interpreted primarily in terms of relative parameter rankings rather than exact variance shares. Several first-order indices are negative and some total-order indices exceed one, indicating Monte Carlo error in the estimator and the presence of strong interaction effects. Consequently, the analysis is most informative for identifying which parameters exert the greatest influence on model outcomes rather than for attributing precise proportions of variance. 

An interesting contrast emerges when comparing the Sobol results with the Latin Hypercube screening analysis. The Latin Hypercube results suggested that mobility outcomes were driven primarily by behavioural thresholds and social-capital dynamics, with parametes such as the capability threshold ($\theta$), social-capital decay ($\gamma_S$), and diaspora influence ($\beta_D$) exhibiting the strongest monotonic associations with model outcomes. By contrast, the Sobol analysis assigns a subtantially larger role to the baseline shock arrival rate ($\lambda_0$), which ranks among the most influential parameter for several outcomes once non-linear effects and parameter interactions are taken into account. This difference suggests that climate exposure exerts much of its influence indirectly through interactions with behavioural and social processes. In other words, thresholds and social capital dynamics remain the most visible drivers in a screening analysis, but the variance-based decomposition reveals that underlying climate exposure becomes equally important once feedback and interactions are considered. This highlights the value of combining screening and variance-based sensitivity methods, as they capture different aspects of model behaviour. 

\newpage
\bibliographystyle{authordate2}

\end{document}